# A Model-Driven Approach to Reengineering Processes in Cloud Computing

**Mahdi Fahmideh**
University of Southern Queensland, Australia

**John Grundy**
Monash University, Australia

**Ghassan Beydoun**
University of Technology Sydney, Australia

**Didar Zowghi**
University of Technology Sydney, Australia

**Willy Susilo**
University of Wollongong, Australia

**Davoud Mougouei**
University of Southern Queensland, Australia

**Context.** The reengineering process of large data-intensive legacy software applications ("legacy applications" for brevity) to cloud platforms involves different interrelated activities. These activities are related to planning, architecture design, re-hosting/lift-shift, code refactoring, and other related ones. In this regard, the cloud computing literature has seen the emergence of different methods with a disparate point of view of the same underlying legacy application reengineering process to cloud platforms. As such, the effective interoperability and tailoring of these methods become problematic due to the lack of integrated and consistent standard models.

**Objective.** We design, implement, and evaluate a novel framework called *MLSAC (Migration of Legacy Software Applications to the Cloud)*. The core aim of MLSAC is to facilitate the sharing and tailoring of reengineering methods for migrating legacy applications to cloud platforms. MLSAC achieves this by using a collection of coherent and empirically tested cloud-specific method fragments from the literature and practice. A metamodel (or meta-method) together with corresponding instantiation guidelines is developed from this collection. The metamodel can also be used to create and maintain bespoke reengineering methods in a given scenario of reengineering to cloud platforms.

**Approach.** MLSAC is underpinned by a metamodeling approach that acts as a representational layer to express reengineering methods. The design and evaluation of MLSAC are informed by the guidelines from the design science research approach.

**Results.** Our framework is an accessible guide of what legacy-to-cloud reengineering methods can look like. The efficacy of the framework is demonstrated by modeling real-world reengineering scenarios and obtaining user feedback. Our findings show that the framework provides a fully-fledged domain-specific, yet platform-independent, foundation for the semi-automated representing, maintaining, sharing, and tailoring reengineering methods. MLSAC contributes to the state of the art of cloud computing and model-driven software engineering literature through (a) providing a collection of mainstream method fragments for incorporate into various scenarios of reengineering processes and (b) enabling a basis for consistent creation, representation, and maintenance of reengineering methods and processes within the cloud computing community.

## 1 Introduction

The reengineering processes for making legacy applications cloud-enabled involve different co-existing and interacting elements such as tasks, procedures, people, resources, and many more [1],[2]. A variety of facets and concepts of those elements transpire such as legacy application code refactoring, interoperability across multiple cloud platforms, architecture design, and an optimized distribution of application components over cloud servers, to name a few [3]. On the other hand, in practice, method engineers who oversee such processes need to know many of these concepts, however, in practice, they apply only an appropriate subset to an ongoing reengineering project [4],[5],[6]. If method engineers are newcomers to the cloud computing field, it may not always be clear what exact tasks and responsibilities should be performed before, during, and after migrating legacy applications to the cloud. The complexity of such a transition and accounts of breakdowns in cloud migration projects have been highlighted by many examples [7],[8]. Indeed, some IT-based organizations have even been unfortunate and forced to move back their cloud-enabled applications to on premises, i.e., de-migrated, after they failed to attain anticipated goals [9]. Among others, failures are often rooted in the lack of timely expertise, inapplicability, and negligibility of past reengineering experience. Former cloud migration experiences are sometimes deemed general, limited to legacy application type/domain, specific to a cloud platform provider, e.g., Amazon, IBM, Cisco, and confined to a particular type of service delivery model, e.g., IaaS (Infrastructure as Service), PaaS (Platform as a Service), and SaaS (Software as a Service) [3]. Moreover, the requirements of a migration context may be quite project-specific and heterogeneous in terms of the choice of service delivery models, security, and scalability. Naturally, method engineers may find off-the-shelf reengineering methods individually incomplete in supporting the overall reengineering process or they may encounter the issue of nonconformity among these methods due to competing requirements and their different viewpoints on reengineering processes.

We refute the suggestion that the extant reengineering methods (e.g., [3],[10]) for cloud migration are not suitable or individually selectable, however, we benefit from a synergistic combination of these methods. Instead of proposing a new reengineering method or aiming for designing a comprehensive and universal one to be applicable to all reengineering scenarios, which is likely infeasible [11],[12],[10], we advocate the development of a foundational middle knowledge layer that pulls together various dispersed and ad-hoc methods describing reengineering processes to cloud platforms. As we will discuss, this view looks outward and claims that cloud-specific reengineering methods exhibit similar underlying concepts and axiomatic commonalities, though they vary in execution details such as the choice of cloud platforms and expressed terminologies. However, such conceptual links have not yet been exposed nor fully exploited to enable extensible and tailorable methods for a legacy to cloud migration, despite the need shown by the earlier research [3],[13],[14, 15].

Against this backdrop, we leverage a model-driven software engineering (MDSE) approach [16],[17]. We use *metamodeling* [18],[19], a particular component of MDSE [20], that is used to model, integrate, and maintain the different software engineering methods (or methodologies) [18],[21],[22]. This has been an encouraging factor for this research to cumulatively build on the prior metamodeling research [23] and to develop a new metamodel specific to legacy application reengineering to the cloud.

Our proposed framework, is called *Migration of Legacy Software Application to the Cloud* (MLSAC), extends our earlier work in [13] and [14] by providing (i) an improved version of the metamodel including new method fragments to provide end-to-end coverage of reengineering lifecycle process, (ii) a tailoring procedure and model-transformation rules to instantiate the metamodel to represent situation-specific reengineering methods, and (iii) implementation of the metamodel, i.e. software tool, to put the framework into real-world applications of situational method creation and maintenance. MLSAC provides a unified view of reengineering methods with the following benefits:

(i) providing a collection of pre-made and reusable method fragments (or process fragments), organized into generic models that allow creating cloud-specific bespoke reengineering methods or at least changes to existing (in-house or off-shelf) methods; and

(ii) facilitating communication among software teams, consistent maintenance, and interoperability of evolving reengineering methods.

Our MLSAC idea, which is in line with the *separation of concerns design principle* [24] in conventional software engineering and an *a la carte* selection and tailoring practice in (situational) method engineering [25], is cloud platform agnostic. It enables method engineers to concentrate on the method design and leave the method operationalization and variations for a particular scenario open to the software team's decision [18],[26].

We applied the guidelines in Design Science Research (DSR) approach [27],[28] to design, implement, and evaluate MLSAC artefacts. DSR enables researchers to engage problems related to the vanguard of the market. Using the DSR approach, we show the expressive power of the MLSAC, as a language infrastructure, in different real-world reengineering scenarios such as EclipseSCADA in Australia and Hackystat SensorBase in the US. Additionally, we discuss the application of MLSAC in a range of reengineering scenarios by our industry partners in Australia. The evaluation results confirm the merits of our framework in a practical context. These also highlighted further research opportunities.

The paper is laid out as follows. Section 2 presents a reengineering scenario showing the key motivation of this research. This is followed by a discussion on the background of MDSE and metamodeling underpinning the theoretical foundation for the proposed framework in Section 3. Section 4 delineates the design of the framework in line with the guidelines to conduct our DSR approach. The application of MLSAC framework in a three-step evaluation is discussed in Section 5. Related works are presented and compared to our work in Section 6. Finally, after the discussion on the threats to the validity of this research in Section 7, this paper ends in Section 8 where the ways for furthering this research are explained.

## 2 Motivating Scenario

Imagine an exemplar scenario of a cloud migration project EclipseSCADA [29], a supervisory control and data acquisition legacy application that was moved to a private IaaS cloud named NeCTAR in Melbourne, Australia. EclipseSCADA is a type of Internet of Things (IoT) based system [30],[31] allowing administrators to monitor an industrial system remotely via sensors and actuators. The cost of application maintenance was relatively high as it was running on dedicated in-house platforms. The top-level management was interested in moving EclipseSCADA workload from in-house hosted servers to flexible and cheaper models offered by cloud services. EclipseSCADA's components were planned to be deployed across Melbourne and Tasmania NeCTAR cloud regions, i.e., hardware components such as sensors deployed in Tasmania servers and software components were hosted in Melbourne servers. A method engineer took the responsibility to rule out a base method to guide the software team for refactoring and re-hosting EclipseSCADA components in the cloud. Such a method could ensure the consistency of reengineering tasks and the software team's outputs. The method could be enacted by the team to enable EclipseSCADA to utilize NeCTAR cloud services. The developers could enact this base method and choose tools, development libraries, and implementation techniques to operationalize the method. The design of such a base method that could ensure a safe EclipseSCADA reengineering over multiple iterations was a challenging exercise due to issues such as:

*Challenge 1*. The design of an overall base reengineering method would need knowledge about several aspects such as understanding legacy application architecture, creating a new architecture model based on NeCTAR cloud, resource scaling, code refactoring according to NeCTAR, and so on. Unfortunately, the domain knowledge about cloud migration is dispersed in the (multi-vocal) literature and, in some cases, it was incompatible or specific to cloud platforms, which is not reusable to design a new method for this project.

*Challenge 2*. The *one-size-fits-all* assumption to design a super engineering method was not a realistic option. Rather, a tailorable reengineering method to unanticipated requirements of EclipseSCADA project would be needed and updated if new requirements arise.

Assessing the merits and demerits of existing reengineering methods and choose the superior one may not be a feasible practice. Instead, the method engineer can select the suitable fragments from the existing methods and create a situation-specific method to guide this reengineering scenario. Towards addressing these challenges, we offer MLSAC framework that provides a metamodel comprising of critical elements of reengineering processes and thus enables the method engineer to reuse, tailor, and extend this metamodel to the variant of migration types such as re-hosting (lift and shift), moving from an existing cloud platform to another platform, or de-migration from the cloud to in-house platforms.

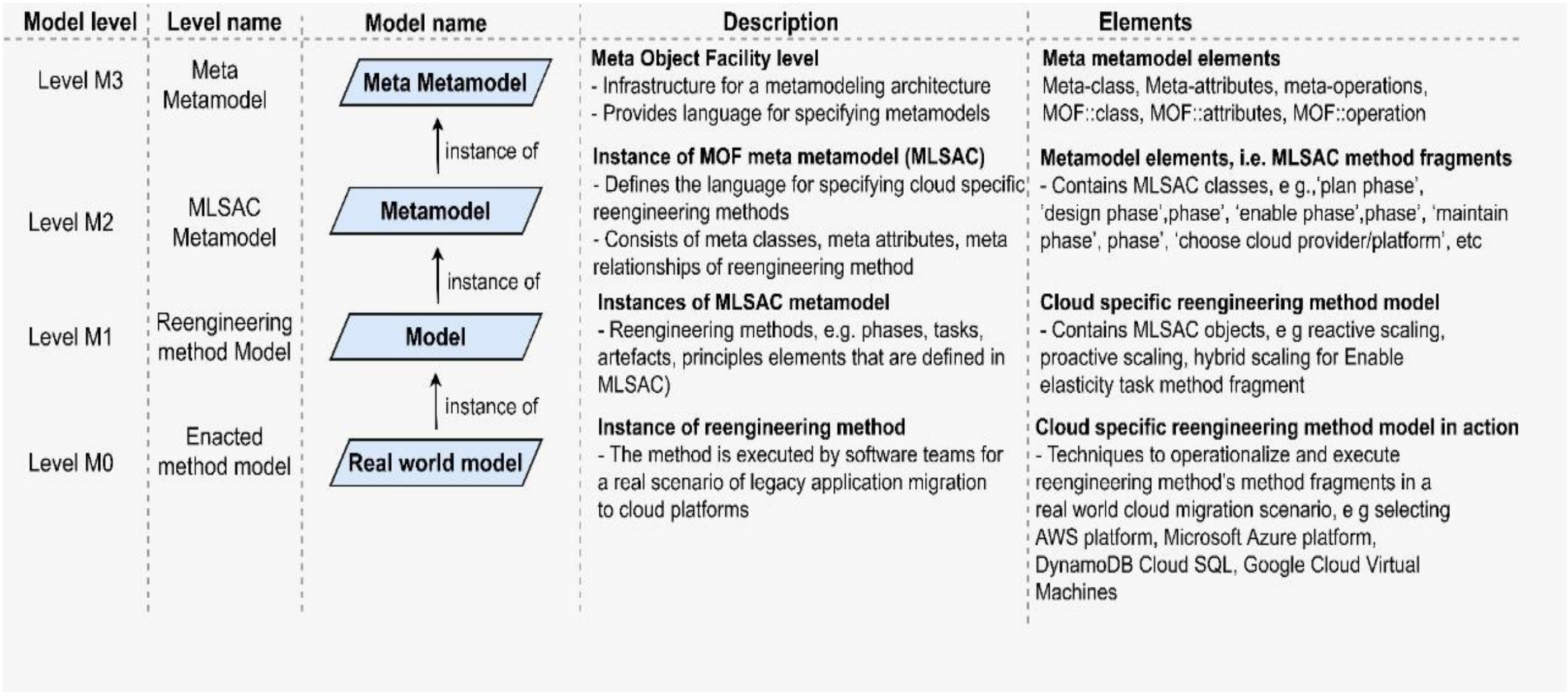

Figure 1. OMG four abstraction levels of hierarchy

# 3 Research Background

A key underlying principle of MDSE approach is the *abstraction*, making a separation between different essential and non-essential aspects of a software system during its design, development, and maintenance [16],[17]. One solution to reach abstraction is to use models. A *model* is a high-level representation of a domain [32]. It is used to manage the complexity of representation and facilitate understanding of the domain for people. Models express the structure and behaviour of the concepts in a domain. Towards this, models may include a set of implying statements and constraints about the concepts and their relationships [33]. Concepts characterise the domain and relationships describe links among these concepts. In MDSE, the term *metamodel* is, evidently, a qualified variant of models. The *metamodeling*, is the act of creating a metamodel, a unified view of a fairly related set of variant models [18].

In this spirit, a metamodel (or meta-method) of reengineering methods specifies phases, tasks, and other elements that should be sequenced and performed in the reengineering process of legacy applications to cloud platforms [32]. Such a metamodel conforming to rules and a defined syntax allow method engineers to construct and maintain customised reengineering methods. Nevertheless, creating such a metamodel, i.e., metamodeling, required, first and foremost, to identify necessary reengineering method fragments. A *method fragment* is a structured, atomic, and re-usable piece of software development processes/methods [34]. In this research, the method fragments are stored in MLSAC repository, combined to make a new method, and matched as reengineering contexts vary and demand.

We leverage the metamodeling foundation proposed by Object Management Group (OMG) [32]. As shown in Figure 1, these layers, which have been used to develop core technologies (e.g. unified modeling language (UML)), define an instance-of-relationship as follow [32]:

(i) M3-level (the meta-metamodel layer or meta-object facility layer) is used to describe basic modeling constructs and their relationships;

(ii) M2-level (metamodel layer) defines concepts and relationships that are instances of concepts from M3 and they define a modeling language to enable model creation/edition at M1;

(iii) M1-level (model layer) instances of M2-level concepts that are used to describe a domain and provide an abstraction of M0-level user data; and

(iv) M0-level is an instance of M1-level, which describes actual user data in a domain model instance.

Each level of the OMG hierarchy (Figure 1) provides a language to express abstractions and relations of concepts at the lower level. The derivation of a model, including its concepts and relations, from its upper level, is referred to as *instantiation* [35]. Based on this modeling hierarchy, MLSAC is placed at M2-level (Figure 1), i.e., metamodel level aiming at the representation of reengineering methods that are situated at M1-level. The methods that are enacted by software teams to perform reengineering scenarios are called method instance, a.k.a. endeavor, and they are positioned at M0-level. The relationship between MLSAC metamodel and method model are defined via the model transformation rules, which converts one model, i.e., source model, to another model, i.e., target model, and thus enables the instantiation of the metamodel to a specific method [36],[37],[38].

# 4 Design Science Research Approach

We employed the DSR approach [27],[28] for the design and evaluation of MLSAC. The DSR aims at rigorously and systematically developing new IT artefacts such as constructs, models, methods, frameworks, and instantiations to address a problem of high significance for research and practice. In this paper, the IT artefact in the focus of the DSR approach is MLSAC framework, which intends to support method engineers in creating and reusing project-specific reengineering methods for cloud migration. To organize this research effort, we used the typical DSR phases of *design* (subsection 4.1,4.2,4.3,4.4) and *evaluation* (Section 5) as described next.

## 4.1 Modeling quality factors and requirements

Following the DSR approach, any novel IT artifact should be designed and evaluated with respect to its pursuit goals. In line with challenges 1 and 2 listed in Section 2, we leverage three general semiotic quality factors (or design principles), namely *semantic quality*, *tailorability*, and *pragmatic quality*, proposed by Lindland et al. [39], to design and evaluate MLSAC. The basic assumption of these factors is that a model is expressed in some language to represent some domain and has some audience. In the context of this research these factors are defined as follows:

*(i) Semantic quality* is the extent to which a model is sound and complete in capturing domain concepts [39]. Through the identification of frequently used concepts in a domain, it will be likely that the resultant model is generic and inclusive. Defining a threshold for model completeness depends on the application context and modeling purpose. We leveraged the highlighted challenges in migrating legacy applications to cloud platforms (e.g., [1],[4]), as a yardstick to derive an initial set of method fragments and their relationships. These challenges are related, for instance, to resource elasticity, multi-tenancy, multiple-cloud platform interoperability, application licensing, dynamicity and unpredictability, and legal issues. The semantic quality is primarily associated with challenge 1.

*(ii) Tailorability quality* is the extent to which a model can be specialized to the fit requirements of a particular domain modeling [39]. Undoubtedly, different reengineering scenarios entail different methods. For example, necessary method fragments that are needed for incorporation into a reengineering process of moving large and distributed workloads from on-premises data centres to public IaaS may vary compared to a reengineering process to enable a legacy application serving as a SaaS. The tailorability factor addresses challenge 2.

*(iii) Pragmatic quality* is the extent to which a model is perceived to be applicable by its audience [39] in terms of properties such as clear and unambiguous diagrams, notations, visualization of relationships, layout, etc. This quality factor concomitantly addresses both *challenges 1* and *2*.

## 4.2 Metamodeling

Our metamodel derivation is based on our earlier conceptual modeling endeavour and uses top-down and bottom-up steps [1],[13],[14],[40]. That is, we used top-down steps to review the general cloud migration literature to get a broad understanding of legacy application reengineering processes to cloud platforms. We also used bottom-up steps for analyzing, reconciling, and abstracting frequently occurring method fragments from the literature. The metamodeling endeavor, the main concern in [13],[14], was iterative and it consisted of the following steps:

*(i) Preparing knowledge source.* This step identified the knowledge source as the input for the metamodeling effort. We utilized the cloud computing literature as the main knowledge source. A major role in this step was played by Kitchenham et al. guidelines for conducting Systematic

Literature Review (SLR) [41] of cloud migration research. The criteria included (a) time filter selecting papers between 2007 and 2019, (b) papers scope restricted to those properly describing the adaptation of legacy applications to cloud platforms, (c) focus forum restricted to international Software Engineering or Information Systems related journals/conferences or multi-vocal literature published by leading companies such as Oracle, IBM, and Amazon. The main keywords for the search of mainstream scientific digital libraries such as Google Scholar, IEEE Explore, ACM Digital Library, Elsevier, SpringerLink, and ScienceDirect were *Cloud*, *Cloud Migration*, *Legacy Application, Reengineering, Method*, and *Process Models*. Different search strings were defined using logical operators OR to cover synonyms for each search string as well as the logical operator AND to link together each set of synonyms. We selected studies from the literature that contained well-described validation such as case study, exemplar scenario, purposeful interview, questionnaire survey of domain experts, simulation, comparative analysis, and theoretical evaluation [41], [42]. Based on this criterion, theory, opinions, white, and short papers with any sort of validation were excluded from the identified studies. The choice of this criterion, as it strived to benefit from the empirical knowledge of cloud migration, could contribute to the reliability of the derived metamodel from the literature. This step identified 74 (seventy-four) studies as publicly listed in [43]. We refer to them as the knowledgebase throughout this paper. Each study has a unique index (from [1..74]).

*(ii) Identifying method fragments.* As discussed in [13], [14], we reviewed each paper and extracted the relevant text segments that could be considered as a method fragment. That is, a *task* is a discrete and small unit of migration work that developers execute. A task execution achieves one or more specific goals, and it produces a tangible *work-product.* A *phase* is a logical way to manage and classify tasks and work-product based on their relatedness. On the other hand, some text segments could be labeled as a *design principle*, which is incorporated into the cloud-enabled legacy application architecture design.

We tended in the selection of method fragments that were frequently underlined in the identified studies, sufficiently cloud-platform independent, and relatively applicable to a variety of reengineering scenarios. Method fragments that were too general or belonged to the general software reengineering such as governance and umbrella activities, risk management were omitted as they could make the metamodel too large and change the scope of this research.

*(iii) Harmonizing method fragments.* The variant definitions of the same method fragments were reconciled. Among several definitions of an overarching method fragment, a hybrid one encompassing all variant definitions was chosen. For example, in studies [S4],[S9],[S32] (as listed in [43]) the choice of cloud computing platform has been defined in three ways, though underlying the same logic: "*this step will select the best supplier based on value, sustainability, and quality*" [S4]; as *identify a set of potential cloud computing platforms based on a project's nature, data confidentiality and sensitivity requirements, budget constraints and long-term organizational objectives*" [S9]; and as *selecting appropriate technology for the modernized system and technology that can run alongside and communicate with the legacy system*" [S32]. A hybrid definition that could cover all these interpretations was chosen for the method fragment *Choose cloud platform/provider* as "*Define a set of suitability criteria that characterize desirable features of cloud platforms. The criteria include provider profile (pricing model, constraints, offered QoS, electricity costs, power, and cooling costs), organization migration characteristics (migration goals, available budget), and application requirements. Based on the criteria identify and select suitable cloud providers*".

*(iv) Organizing method fragments into phases.* Reconciled method fragments in the previous step were grouped into one

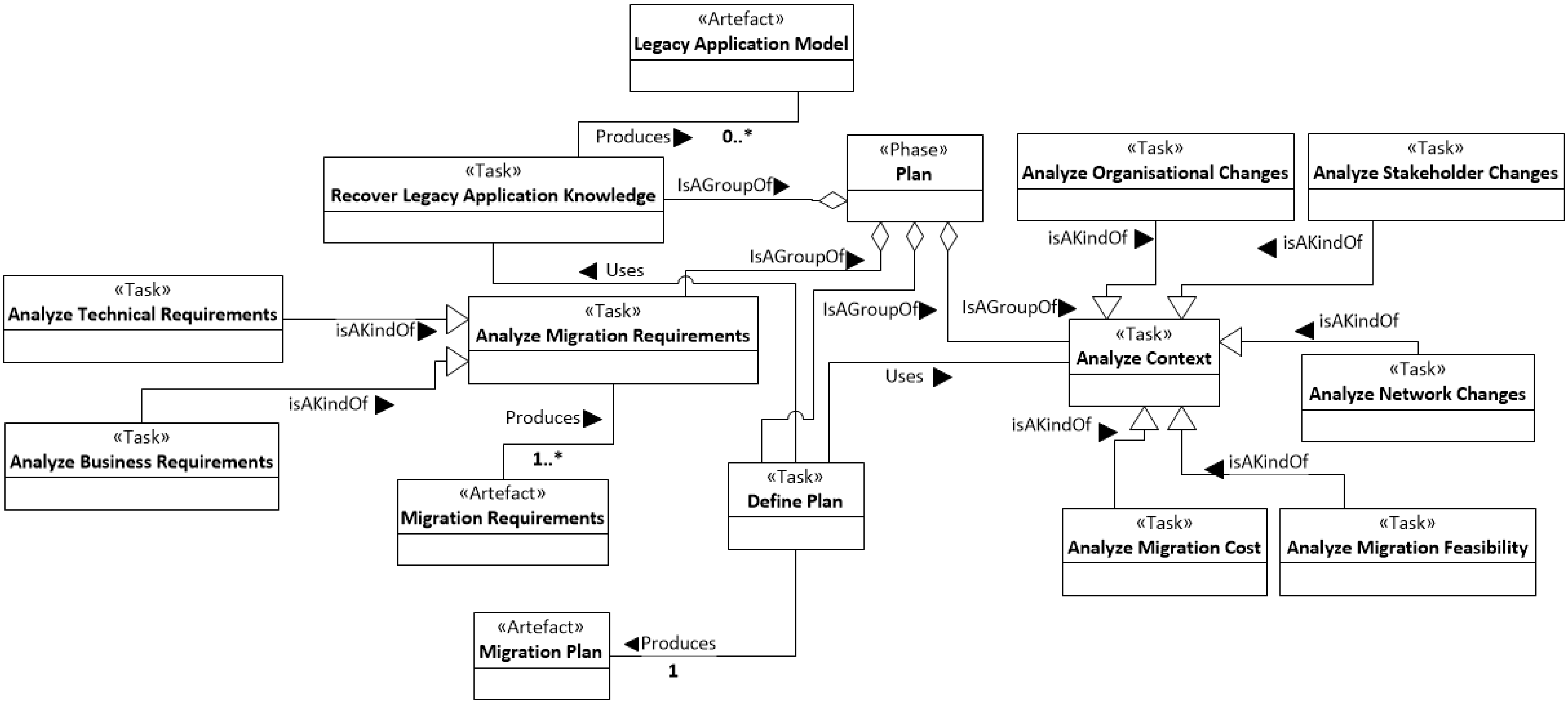

Figure 2.a Method fragments of *Plan* phase

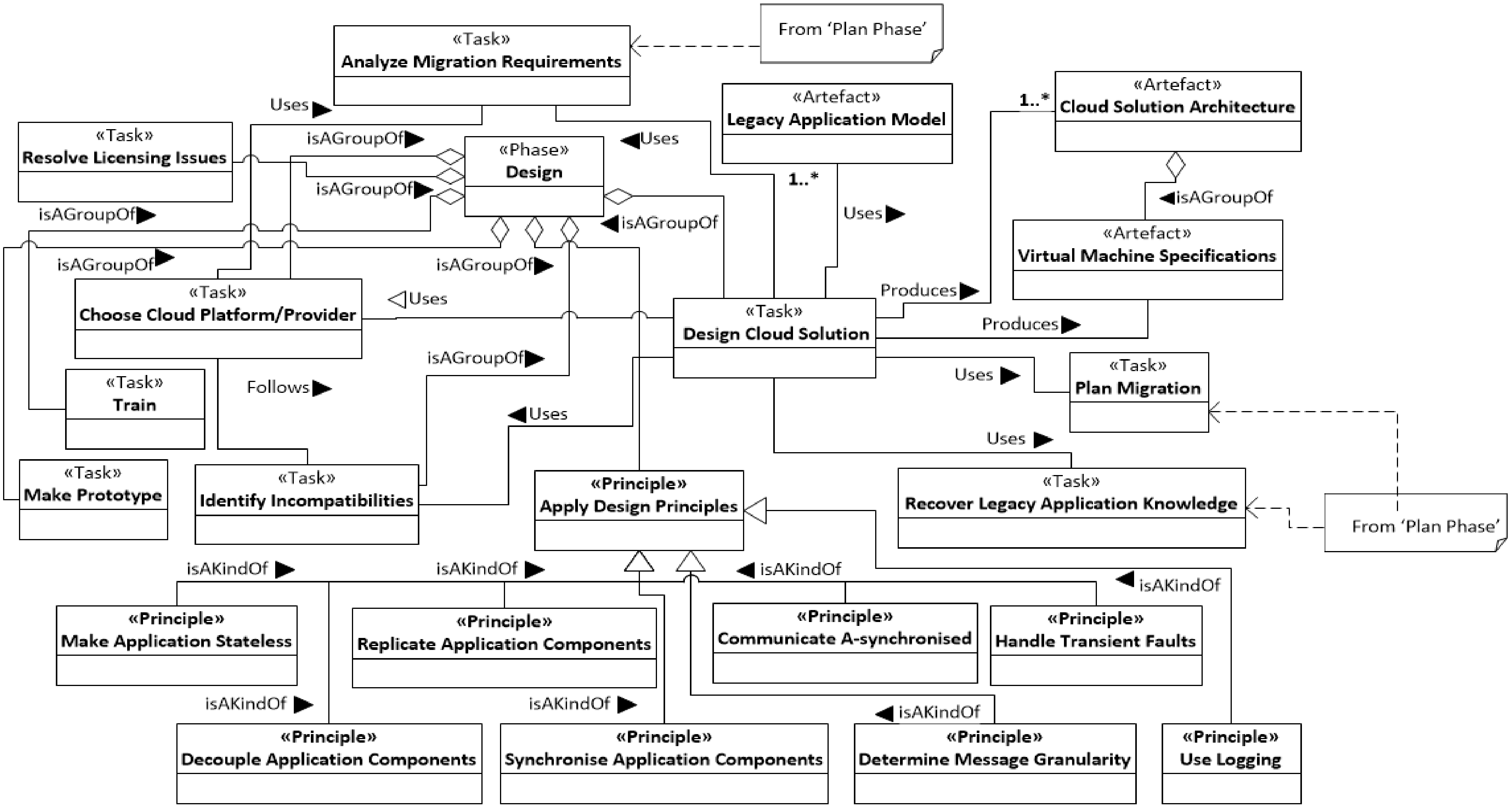

Figure 2.b Method fragments of *Design* phase

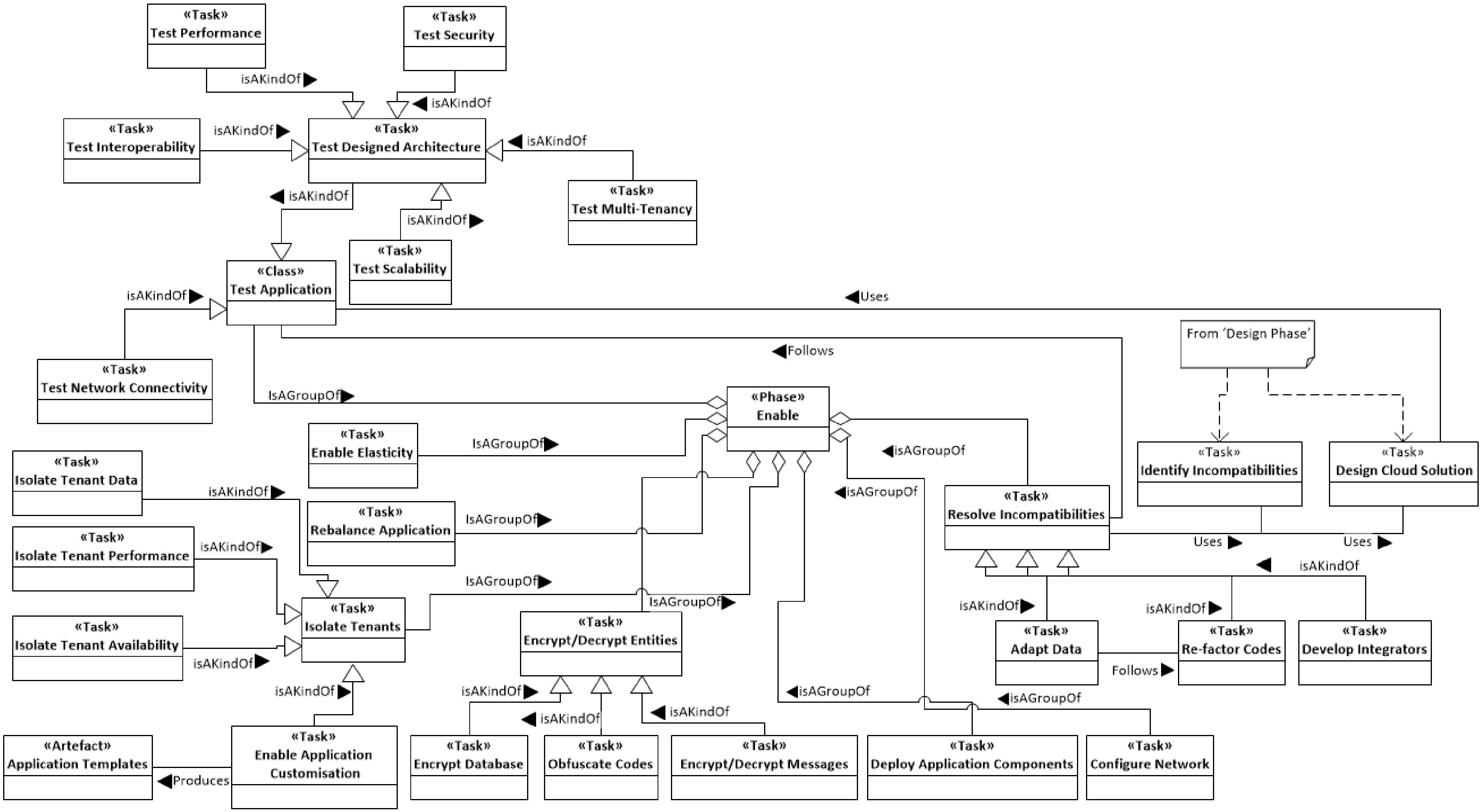

Figure 2.c Method fragments of *Enable* phase

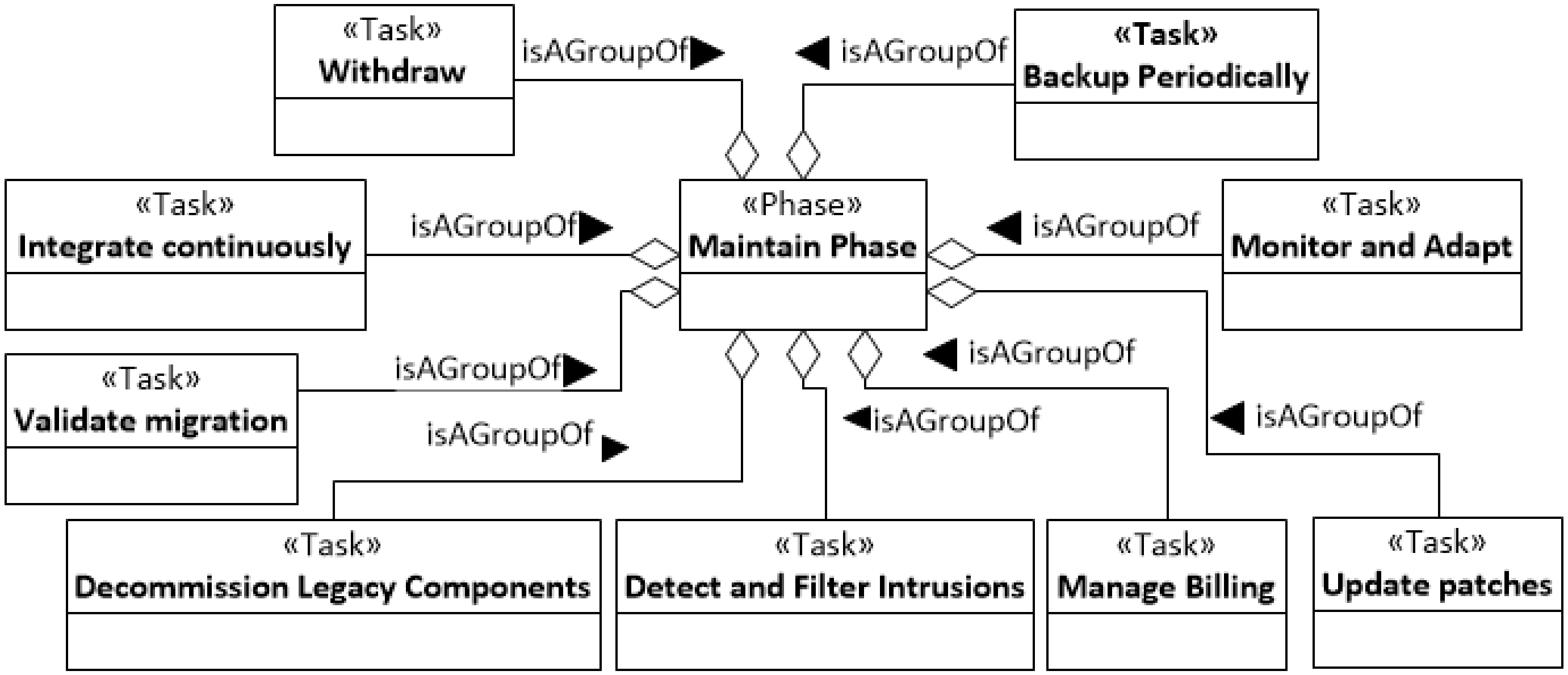

Figure 2.d Method fragments of *Maintain* phase

Table 1. An excerpt of transformation rules for instantiation of MLSAC method fragments to a reengineering process

| Rule id | Rule name | Rule meaning | Rule syntax |
|---|---|---|---|
| R00 | *ResolveIncompatibilities* method fragment of MLSAC | ResolveIncompatibilities::=(< ResolveIncompatibilities MethodFragmentClass>) | $C_M(\mu MLSAC)$ |
| R01 | Instance model of MLSAC | The set of all MLSAC metamodel instances, i.e., methods, conforming to MLSAC metamodel, $\mu MLSAC$ | $\top(\mu MLSAC)$:={: $(\mu MLSAC) \mid\equiv \mu MLSAC$ }8 |
| R01.1 | MLSAC metamodel fragment | MLSAC metamodel method fragment is consisting of name and relationships with other method fragments such as sequence, association, specialization, and aggregation | $C_M(\mu MLSAC) ::= MLSAC_MethodFragment$); MLSAC_MethodFragment ::=(<MethodFragmentName> AND <MethodFragmentRelationship) |
| R04 | Method fragment subset MLSAC metamodel | All method fragments defined in each class of phases are a subset of MLSAC metamodel | $C_M(\mu MLSAC_PlanPhase)$^ $C_M(\mu MLSAC_DesignPhase)$ ^ $C_M(\mu MLSAC)_EnablePhase$ ^ $C_M(\mu MLSAC_MaintainPhase)$ |
| R04.3 | *Plan phase* M1 method | *Plan phase* of a method, which contains method fragments and their relations from Plan phase class of MLSAC metamodel, is a part of a design method | $(C_M(\mu MLSAC_PlanPhase)$ ^ r($\mu MLSAC_PlanPhase$)) ∊ : $\top(\mu MLSAC_PlanPhase)$ |
| R05.1 | Relationships of method fragments | Relationships of all method fragments, as defined in each class of phases, are a subset of MLSAC metamodel | $(r(C_M(\mu MLSAC_{PlanPhase}))$^ $r(C_M(\mu MLSAC_DesignPhase))$ ^ $r(C_M(\mu MLSAC)_EnablePhase))$ ^ $r(C_M(\mu MLSAC_{MaintainPhase})) \subset \mu MLSAC)$ |

of the generic reengineering phases: *Plan*, *Design*, *Enable*, and *Maintain*. The *Plan* phase is to understand the organizational context in which legacy applications operate. The *Design* phase defines a new cloud-enabled architecture for the legacy applications. The *Enable* phase lists the tasks such as incompatibility resolutions and network configuration to implement the cloud-enabled architecture that is defined in Design phase. The *maintain* phase deploys and monitors the performance of application components running over cloud platforms.

*(v) Conceptual representation.* Our metamodeling effort resulted in several UML object models, as shown in figures 2.a, 2.b, 2.c, and 2.d. These include key relationships, associations' cardinalities, and stereotypes such as phase, task, and work-product. Together, they constitute a set of commonly occurring method fragments for incorporating into a typical reengineering method. The operationalization of the method fragments is deferred to implementation time and subjected to developers' choice of techniques and tools. For example, whilst method fragment *Develop Integrator* (Figure 2.b) informs developers of incorporating mechanisms to address the interoperability and portability of applications, a Docker container, i.e., a form of virtualization technology, can be used to realize this method fragment.

## 4.3 Model transformation rules

The model transformation rules instantiate MLSAC metamodel at M2-level to generate new specific method instances at M1-level according to MOF framework (Fig. 1). This instantiation is a vertical transformation from the higher level of abstraction, i.e., MLSAC metamodel, to the lower-level model, i.e., MLSAC metamodel instance [35]. We defined the transformation rules based on our knowledge source (Appendix A). These rules act as guidelines and provide semantics for the transformation of the metamodel to a specific reengineering method instance. They guarantee a consistent transformation from the metamodel (M2-level) to a model (M1-level) [38]. The transformation rules are implemented as a distinct module in MLSAC architecture using database tables describing relationships among method fragments. The transformation rules, like semi-MOF transformation notations [44], $C_M(\mu MLSAC)$ indicate a set of method fragments ∊ MLSAC metamodel. Table 1 shows a sample of rules. For example, Rule R00 formalizes the instance creation of *Plan* phase in MLSAC to a specific process model where:

> *Transformation rule*: Rule R00 (Plan phase):
> *Rule syntax*: $C_M(\mu MLSACplanphase)$
> *Rule meaning*: The set of tasks defined in MLSAC Plan phase are instantiated to Plan phase of a new method
> *Rule construct*: <*PlanPhaseClass*> :: = (*Analyze business requirements* Class AND *Analyze migration cost* Class AND *Analyze migration feasibility* Class AND Analyze network change Class AND *Analyze organizational changes* Class AND *Analyze stakeholders change* Class AND *Analyze technical requirements* Class AND *Define Plan* Class AND *Recover legacy application knowledge* Class)

Table 2. An excerpt of relationships matrix for selection of method fragments based on migration types (√: Mandatory, (√): Situational, × Unnecessary), from [1]

| Method fragment | Migration type* | | | | | Situation |
|---|---|---|---|---|---|---|
| | I | II | III | IV | V | |
| Adapt data | × | (√) | (√) | (√) | (√) | The incorporation of this fragment for the migration types II, III, IV, V depends on the choice of a cloud platform and inconsistencies between legacy application platform and cloud platform. |
| Analyze business requirements | √ | √ | √ | √ | √ | Mandatory |
| Choose cloud platform/provider | √ | √ | √ | √ | √ | Mandatory |
| Cloud solution architecture | √ | √ | √ | √ | √ | Mandatory |
| Decouple application components | (√) | (√) | (√) | (√) | (√) | The incorporation of this principle depends on new designed architecture model and the distribution of application components in the cloud. |
| Develop integrators | (√) | (√) | (√) | (√) | (√) | The incorporation of this fragment depends on the choice of a cloud platform and required effort to refactor/modify legacy codes. If the code refactoring, as supported by refactor codes, is costly, then developing integrators/adaptors can be served as an alternative solution to hide incompatibilities. |
| Enable elasticity | (√) | (√) | × | × | (√) | The incorporation of this fragment in the migration types I, II, and V depends on a need for the application elasticity. |
| Encrypt/decrypt database | × | (√) | (√) | (√) | (√) | The incorporation of this fragment in the migration types depends on security requirements. |
| Handle transient faults | √ | √ | √ | √ | √ | Mandatory |
| Identify incompatibilities | √ | √ | × | √ | √ | Mandatory |
| Isolate tenant availability | × | √ | × | × | × | This is a mandatory fragment for migration type II. |

*Migration type I: deploying business logic of a legacy application on cloud via IaaS service delivery model, type II: replacing or reengineering legacy components with SaaS delivery model, type III: deploying legacy database components on cloud data storages, type IV: converting legacy database components to cloud database solutions, and type V: deploying whole legacy application stack on cloud via IaaS service delivery model

Each cloud migration type, such as I, II, III, IV, and V [3], entails the incorporation of some specific tasks into the reengineering process. To this end, the relationship matrix is used to classify method fragments and to guide the metamodel instantiation. As such, the method engineer is informed of method fragments that are required for inclusion into a reengineering method instance. Table 2 shows a sample of situations in which method fragments are associated with a given migration type. These relationships are based on the knowledge source and are coded as transformation rules in MLSAC. For example, according to the knowledge source, reflected in studies [S41] and [S42] (see [43]), deploying legacy application components on cloud servers via IaaS service delivery model, i.e., migration type V, requires a new architecture model specifying a topology of migrated components and their communication with in-house components. Hence, *Design cloud solution* method fragment is mandatory for inclusion in a newly created reengineering method in all migration types. To ensure this, *Rule R01* is defined:

*Transformation rule*: Rule R01 (Plan, Design, Enable, Maintain phases):
*Rule syntax*: $C_M(\mu MLSACplanphase)$ ^ $C_M(\mu MLSACdesignphase)$ ^ $C_M(\mu MLSACdesign\ phase)$ ^ $C_M(\mu MLSACmaintainphase)$
*Rule meaning*: The set of mandatory task method fragments defined in MLSAC *Plan* phase is instantiated to all phases of a new reengineering process

*Rule syntax*: <*PlanPhaseClass*> :: = (*Plan phase* Class AND *Design phase* Class AND *Enable phase* Class AND *Maintain phase* Class AND *Analyze business requirements* Class AND Analyze migration cost Class AND *Analyze migration feasibility* Class AND *Analyze network change* Class AND *Analyze organizational changes* Class AND *Analyze stakeholders change* Class AND *Analyze technical requirements* Class AND *Choose cloud provider* Class AND *Cloud solution architecture model* Class AND *Define plan* Class AND *Deploy application components* Class AND *Handle transient fault* Class AND *Legacy application architecture* Class AND *Migration plan* Class AND *Migration requirements* Class AND *Synchronize application components* Class AND *Test network connectivity* Class AND Test security Class)

The relationships between method fragments (figures 2.a, 2.b, 2.c, and 2.d) such as *follows*, *association*, *specialization* and, and *aggregation*, respectively, are represented by notations (−), (—▷) and (—◇) . They have been defined in MLSAC based on the knowledge source and metamodeling steps. They are stored in MLSAC repository and are used during MLSAC metamodel instantiation and tailoring. For example, *follows relation* means that the default execution sequence of method fragments in a typical migration scenario. As shown in Table 3, the relation between the method fragments *Identify incompatibilities* and *Choose cloud platform* signifies that once a cloud platform is chosen, examining potential incompatibilities between legacy application components and this platform should

Table 3. Sample relationships between the method fragments in the metamodel (L: knowledge source [43] M: during metamodeling steps)

| Relationship Type | Relationship sub-type | Method fragment 1 | | Method fragment 2 | | Source |
|---|---|---|---|---|---|---|
| | | Name | Type | Name | Type | |
| Association | Uses | Analyze migration requirements | Task | Choose cloud provider | Task | L |
| Association | Uses | Design cloud solution | Task | Plan migration | Task | L |
| Association | Follows | Choose cloud provider | Task | Identify incompatibilities | Task | L |
| Association | Produces | Design cloud solution | Task | Cloud solution architecture | Work-Product | L |
| Aggregation | IsAGroupOf | Analyze context | Task | Plan | Phase | M |
| Specialization | isAKindOf | Re-factor codes | Task | Resolve incompatibilities | Task | M |
| Specialization | isAKindO' | Develop integrators | Task | Resolve incompatibilities | Task | M |
| Specialization | isAKindOf | Adapt data | Task | Resolve incompatibilities | Task | M |

be the next task. The evidence to this, based on the knowledge source, is:

> "*An application is analyzed to assess its compatibility with the potential cloud computing environment. For example, it may be the case that a target PaaS cloud does not support frameworks or specific technologies being used by an application. If such issues are identified, then these need to be resolved first*" [S36].

The full list of these recommended relationships is publicly available at GitHub [45]. Arguably, the execution order of method fragments is context-dependent and confined to each individual reengineering scenario. Thus, it is not feasible to capture all possible flows in MLSAC. Moreover, method engineers are not restricted to follow the predefined sequences in MLSAC when creating a new method for their own reengineering scenario.

## 4.4 Architecture overview

We implemented the MLSAC prototype system using technologies Microsoft .Net Framework 2015, C# programming language, and Microsoft Access Database. Feedback collected from our partners helped us test and improve MLSAC which is now publicly available at GitHub [45] and briefly described below. Figure 3 depicts a snapshot of MLSAC's three-layer architecture.

The user interface layer has 19 interactive forms enabling a method engineer to create, update, and import/export the metamodel and method instances. The business logic layer operationalizes vertical model transformations from MLSAC metamodel. Since we used the object-oriented paradigm to design MLSAC architecture; three levels of MOF, i.e., *metamodel*, *method model*, and *method model instance* were mapped to the notion of *class*, each with its properties and operations in the business logic layer. The main classes in this layer are *method_fragment*, *metamodel*, *method model (reengineering process model)*, *method_instance*, *modeling component*, and *tailoring* (Figure 3). For example, the *method_fragment* class has the following fields: identifier, type, name, definition, relationships, and migration type. An excerpt of that is:

> MethodFragmentClaas::= <Method_fragment_Id, Method_fragment_name, Method_fragment_type, Method_fragment_definition, Method_fragment_relation, Method_fragment_ migration_type>

where:

> *Method_fragment_Id* represents the identifier of the method fragment;
>
> *Method_fragment_name* refers to the name of the method fragment;
>
> *Method_fragment_type* indicates whether the method fragment is phase, task, work-product, and principle;
>
> *Method_fragment_definition* explains the steps to execute and aspects related to the method fragment;
>
> *Method_fragment_relation* specifies the relation to other fragments such as being precedence and successors; and
>
> *Method_fragment_migration_type* defines the situations for which the method fragment is recommended to be sequenced in the reengineering method.

The layer also defines a *modeling component* providing functions for deriving new bespoke method instances from the metamodel. The *tailoring* class implements functions for creating methods. The method engineer can refine a method fragment from M2-level to M0 level through associating it with operationalization techniques as defined by the *technique* method fragments.

The MLSAC is implemented using a relational database management system (RDBMS). A collection of relational tables is used to keep and update method modeling. Regarding MOF framework, a method fragment is stored at three levels of M0, M1, and M2 (Figure 1). That is, MLSAC metamodel, itself, is expressed as a collection of method fragments that are stored in *Metamodel* table. The metamodel enables describing reengineering methods. Instances of MLSAC metamodel that are positioned at M1-level and enacted instances that are positioned at M0-level are, respectively, stored in *Method* and *MethodInstance* tables. Moreover, information about the operationalization of method fragments, i.e., supportive techniques, positioned at M0-level, is stored in *SupportiveTechniques* table. Some further exemplar applications of MLSAC prototype namely (i) creating and configuring a reengineering method via MLSACA metamodel, (ii) migrating a legacy application data layer to cloud (NovaTec case), and (iii) migrating full legacy application stack to the cloud (Accenture case) are available at [45].

# 5 Framework Evaluation

The MLSAC framework aims at providing a language infrastructure to be used for representation, unification, and sharing cloud-specific reengineering methods. We iteratively evaluated and revisited MLSAC in the view of quality factors (Section 4.1) as shown in Table 4 and discussed in what follows.

Table 4. MLSAC evaluation

| Evaluation scenario | Quality factor | | |
|---|---|---|---|
| | Semantic | Tailorability | Pragmatic |
| EclipseSCADA case (section 5.1) | - | √ | - |
| Hackystat case (section 5.2) | - | √ | - |
| User evaluation (section 5.3) | √ | √ | √ |

## 5.1 Maintaining reengineering methods

This evaluation focused on the *tailorability* factor of MLSAC. In the EclipseSCADA exemplar scenario stated in Section 2, the developers aimed at deploying the application stack on the NeCTAR to reach flexible fees based on required computing resources.

**Evaluation procedure.** The method engineer performs the following tailoring procedure to instantiate the metamodel to represent and maintain the EclipseSCADA reengineering method's content.

*Step i.* Three input parameters to MLSAC are method name/description, the choice of migration types, and phases. This scenario is subsumed under the migration type V, i.e. a virtual-machine-based application migration to the NeCTAR cloud IaaS model [3]. The method engineer merely focuses on the *Plan* phase and skips other phases.

*Step ii.* Input parameters are used to inform a vertical transformation from MLSAC's metamodel – as the source model at M2-level – to a new reengineering method instance at M1-level. Relevant method fragments associated with the selected migration type V and the *Plan* phase are retrieved from the MLSAC repository according to the transformation guidelines for inclusion in the new method instance. These fragments are suggestive by the MLSAC repository for inclusion in the base reengineering method: *Analyze context*, *Recover legacy application knowledge*, *Analyze migration requirements*, *Define plan*. This derivation can be codified using the following pseudo-code (if-then expression):

```
Model_Instance Function MetamodelInstantiation (mt,
p)
  {
    QUERY_STRING ← {};//query string to retrieve method
    fragments from the repository
    MIGRATION_TYPE ←  mt; //the choice of migration type
    PHASE  ← p; //the choice of phase
    if (MIGRATION_TYPE == Type I) then
    QUERY_STRING ← "TYPE I"
    else if (MIGRATION_TYPE == Type II) then
    QUERY_STRING ← "TYPE II"
    else if (MIGRATION_TYPE == Type III) then
    QUERY_STRING ← "TYPE III"
    else if (MIGRATION_TYPE == Type IV) then
    QUERY_STRING ← "TYPE IV"
    else if (MIGRATION_TYPE == Type V) then
    QUERY_STRING ← "TYPE V"
    if (PHASE == Plan_Phase then
    QUERY_STRING ← + "Plan_Phase"
    if (PHASE == Design_Phase then
    QUERY_STRING ← + "Design_Phase"
    if (PHASE == Enable_Phase then
    QUERY_STRING ← + "Enable_Phase"
    if (PHASE == Maintain_Phase then
    QUERY_STRING ← + "Maintain_Phase"
    PROCESS_INSTANCE←retrieve (QUERY_STRING)
    return PROCESS_INSTANCE
```

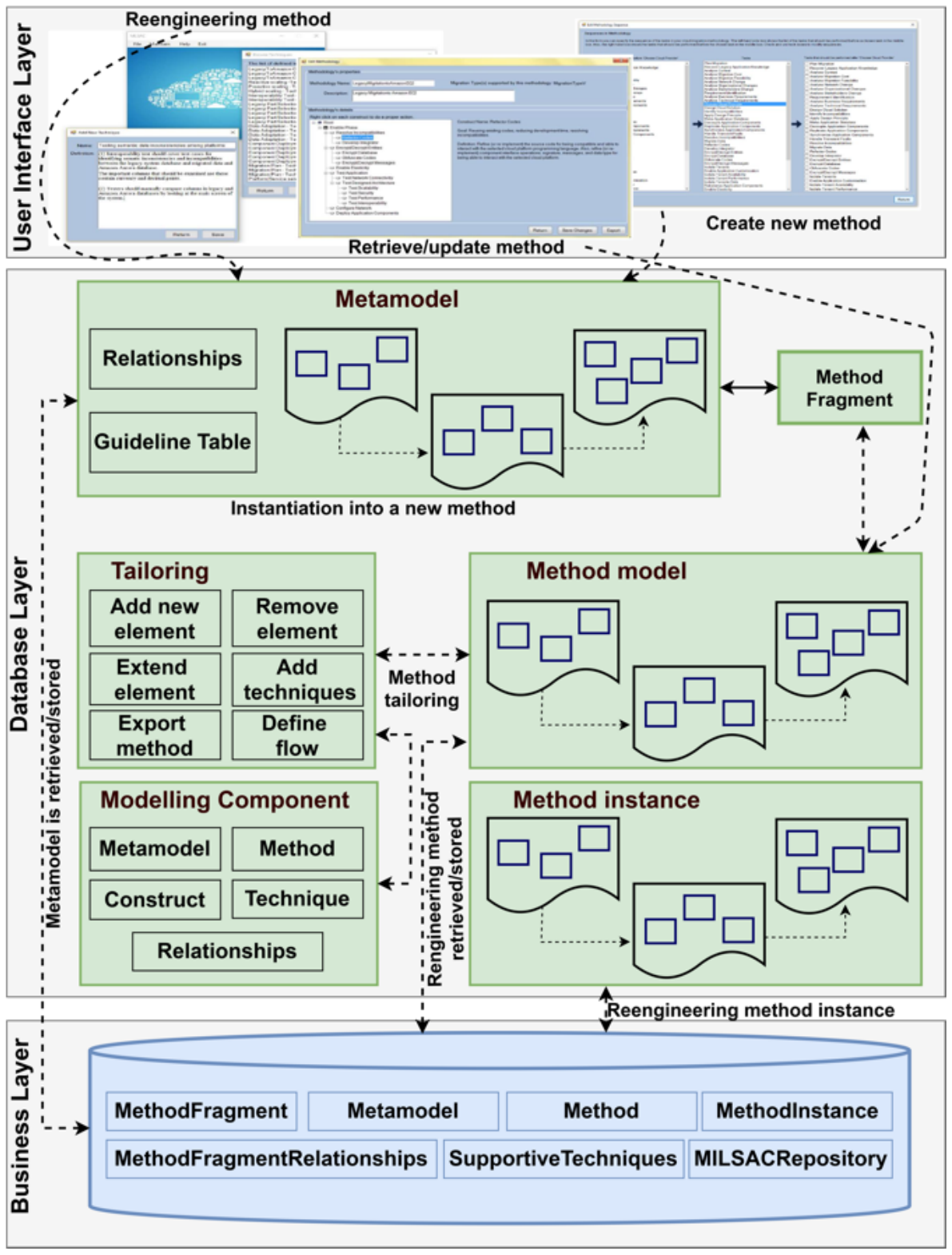

Figure 3. MLSAC overall architecture

}

This method, containing the set of method fragments and their definitions that are reused from the metamodel, is shown in Figure 4.a. The graphical user interface in Figure 4 has three main sections. The upper section has the general description of the

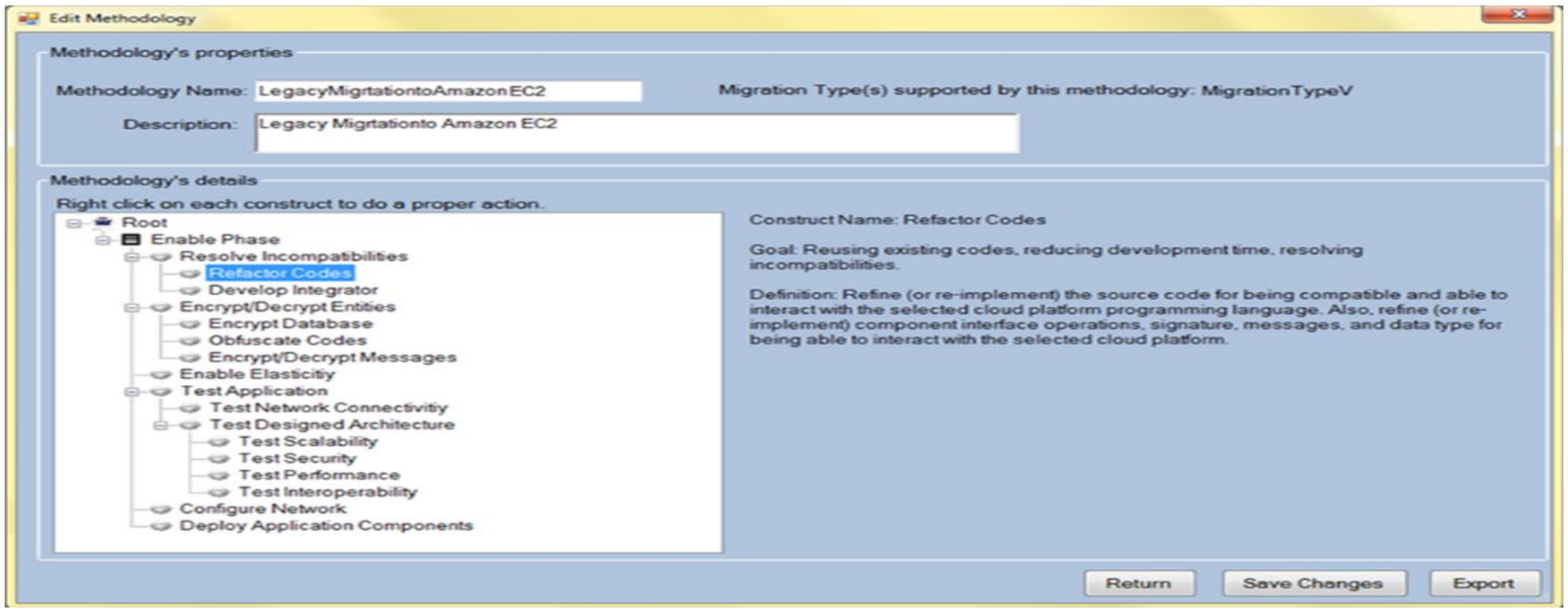

Figure 4.a An instantiated reengineering method from the metamodel

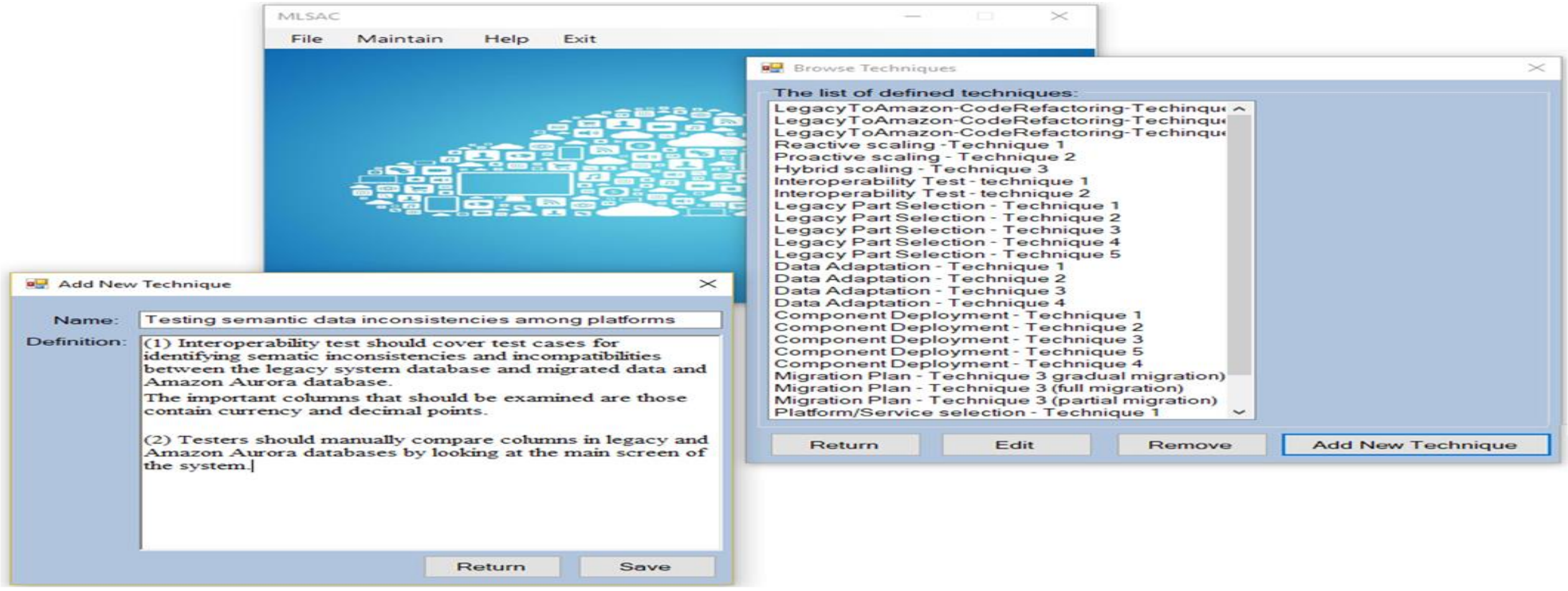

Figure 4.b Defining an implementing technique for *Test interoperability* task method fragment

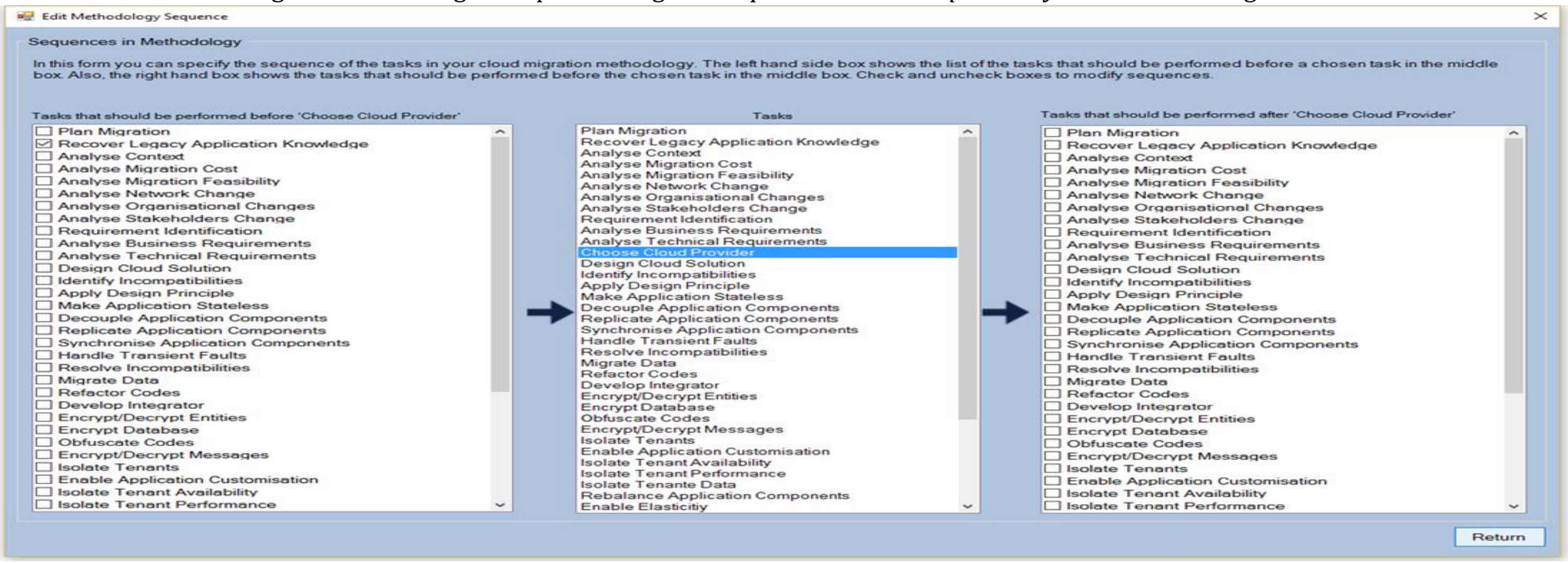

Figure 4.c Specifying sequences among task method fragments

Metamodel data

| Method_fragment |
|---|
| Method_fragment_Id |
| Method_fragment_name |
| Method_fragment_type |
| Method_fragment_definition |
| Method_fragment_relation |
| Method_fragment_ migration_type |

Method fragment schema

| Method_fragment_Id | Method_fragment_name | Method_fragment_type | Method_fragment_definition | Method_fragment_relation | Method_fragment_ migration_type |
|---|---|---|---|---|---|
| MF1 | Enable phase | Phase | The plan phase is to understand the organizational context and legacy applications | Pre:Plan; Nxt: Maintain | Mandatory (I, II, III, IV, V) |
| MF21 | Develop Integrator | Task | Develop a mediator component that resolves mismatches between legacy and cloud services. The mediator component is plugged into the legacy application. It is responsible to wrap/transform incompatibility, e.g., message format, interaction protocols, between the legacy application components and cloud services | Resolve Incompatibilities | Situational (I, II, III, IV, V) |
| … | … | … | … | … | … |

EclipseSCADA_Method (Metamodel instance)

| Method_fragment_Id | Method_fragment_name | Method_fragment_type | Method_fragment_definition | Method_fragment_relation | Method_fragment_ migration_type |
|---|---|---|---|---|---|
| MF1 | EclipseSCADA_ Enable phase | Phase | The plan phase is defined as agreed. | Pre:Plan;Nxt: Maintain | Mandatory |
| MF21 | EclipseSCADA_ Develop Integrator | Task | Use or implement wrappers to resolve incompatibilities occurring between EclipseSCADA and IaaS cloud NeCTAR that are to be plugged to these applications | Resolve Incompatibilities | Mandatory |
| … | … | … | … | … | … |

Task techniques to operationalize task method fragment reengineering method, e.g., name and migration. The bottom-left section shows the method fragments reused from the MLSAC metamodel. The bottom-right section (Figure 4) gives information about a selected method fragment once the method engineer clicks it in the tree view.

| Task_Technique_Id | Task_Technique_Name | Task_Technique_ Definition |
|---|---|---|
| MF24 | Docker container | A form of virtualization technology that can be used to realize this method fragment |
| … | … | … |

| Method_fragment_Id | Task_Technique_Id |
|---|---|
| MF21 | TT12 |
| … | …. |

Figure 5. Example of tables storing the metamodel and its EclipseSCADA instantiation

*Step iii.* The method engineer can perform different optional sub-steps to tailor the base reengineering model instance to meet scenario requirements (Figures 4.b and 4.c). These include (a) adding new method fragments to the method if the pre-existing method fragments in MLSAC repository are insufficient to support requirements of the reengineering scenario, (b) extending the existing method fragments of MLSAC with new ones through the inheritance mechanism, (c) specifying alternative techniques to operationalise the method fragments (Figure 4.b), and (d) defining arbitrary sequences among the method fragments (Figure 4.c). For example, the method engineer defines three custom sub-classes of *Define plan* task method fragments namely *Determine application disposition*, *Plan migration*, and *Define migration road map*. According to OMG modeling framework (Figure 1) [32], the abovementioned sub-steps are horizontal transformations where the method model at M1-level, as the source model, is evolved to its next operational target model to include/exclude method fragments as required in the migration scenario.

*Step iv.* Once tailored, the method model can be either stored in MLSAC database or exported as an XML document to be later shared with developers for the enactment. Figure 5 shows such an instance of the relationships between MLSAC metamodel, *EclipseSCADA_Method*, and *TaskTechniques.*

## 5.2 Creating and tailoring new reengineering methods

MLSAC is generic, and it provides a skeleton for the effective design and maintenance of reengineering methods. Variations to its element operationalization are left open to developers' decisions in a particular reengineering scenario. We examined the tailorability of MLSAC in Hackystat project [46]. This illustrated how an existing reengineering method for an open-source legacy application, called Hackystat SensorBase service, positioned at M1-level, can be created, reused, and maintained via MLSAC instantiation.

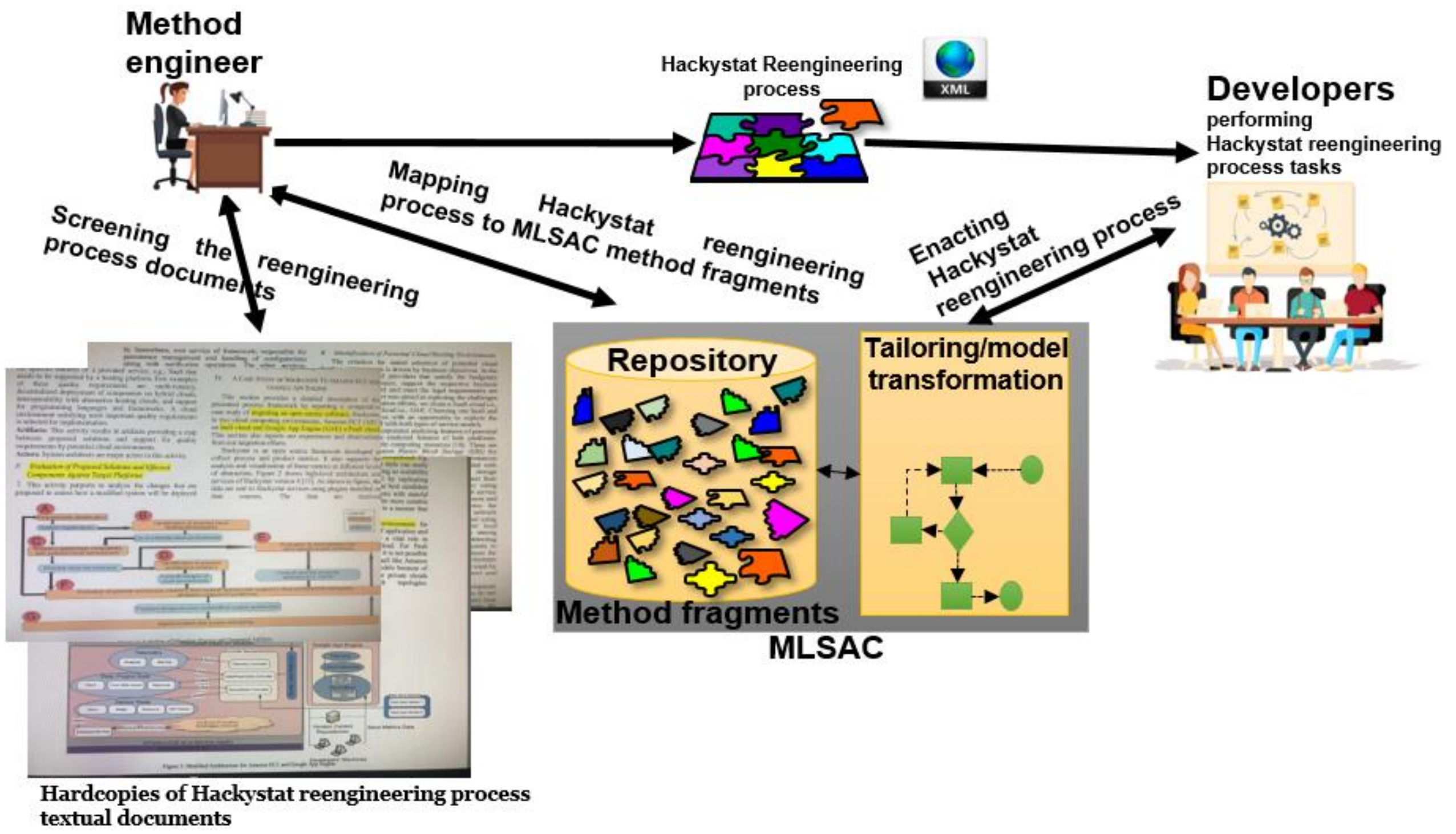

Figure 6.a Hackystat reengineering method document, the entry for MLSAC method fragments

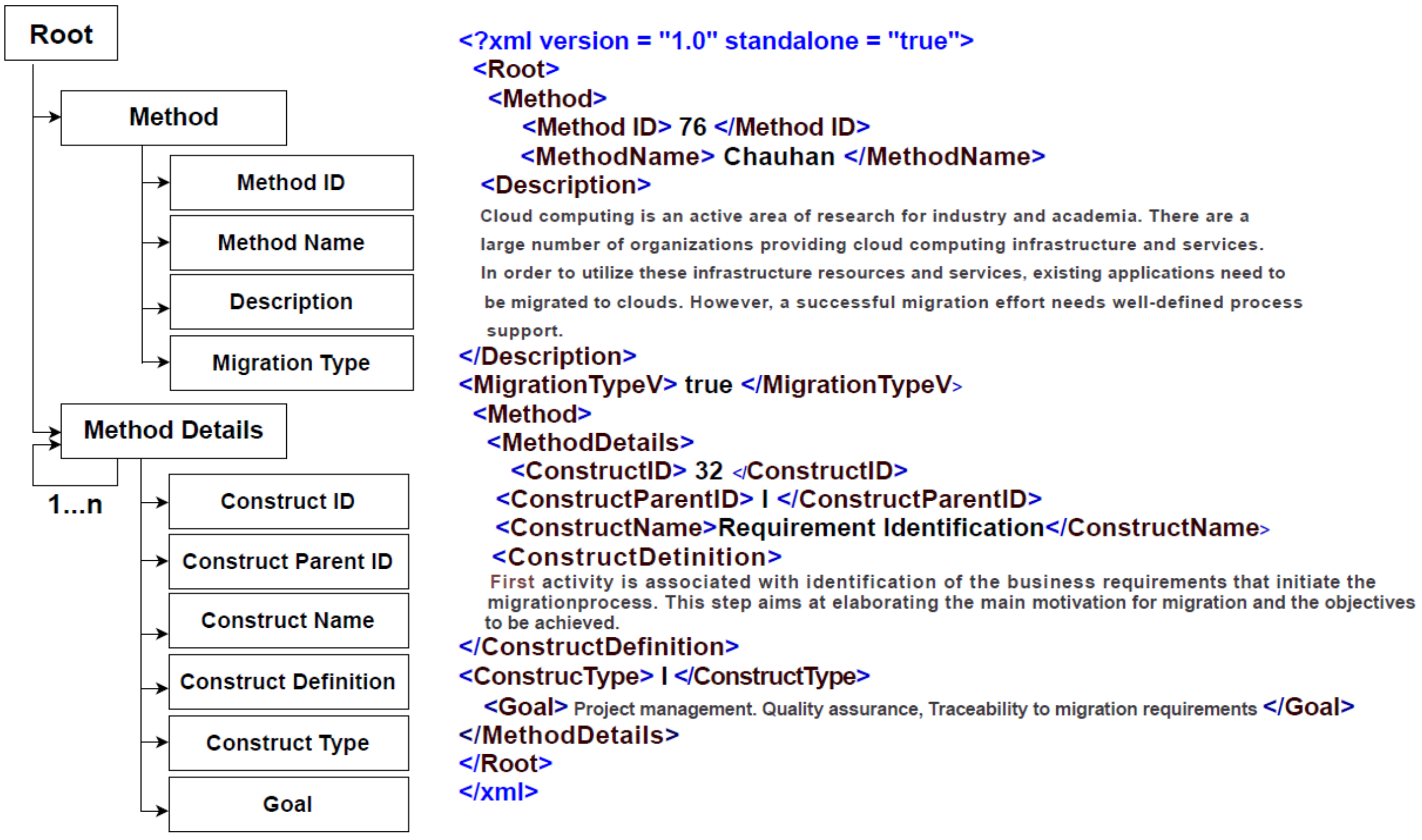

Figure 6.b an excerpt of Hackystat reengineering method stored in MLSAC repository and exported in XML format

The Hackystat method establishes a set of guidelines and practices to be enacted by the software team for migrating Hackystat SensorBase legacy application to SaaS. The method is described in natural language (Figure 6.a). This way of representation is inherently ambiguous and non-modular causes difficulties in the method maintenance. For example, if the method engineer needs to change the Hackystat method content, she should manually update and check the entire method document to ensure the consistency of method content. This is a very time-consuming and error-prone task.

To address these issues, MLSAC enables the method engineer to automatically reuse the method fragments provided by MLSAC to create and tailor the Hackystat method. That is, she can instantiate a SaaS-specific method from MLSAC metamodel including recommended method fragments for incorporation into the reengineering process of Hackystat application to SaaS. She can then customize or extend this method, for example, by adding new method fragments to be sequenced into this reengineering exercise. Furthermore, the method engineer traces the origin of the content, e.g., guidelines and practices, that are reported in the document of Hackystat method to the method. This indicates the extent to which the newly created method preserves the sematic of the Hackystat method.

The above steps are the vertical model transformation according to MOF framework where the method fragments in MLSAC, as the source model positioned at M2-level, are instantiated to represent the Hackystat method, as the target model positioned at M1-level.

Finally, the method is stored in the MLSAC repository. Developers can reuse, customize, and enact this M1-level method for a given scenario at M0-level (Figure 6.a,b).. The Hackystat scenario was:

> *Hackystat is an open-source application developed to collect process and product metrics founded university of Hawaii in the US. It also supports the analysis and visualization of these metrics at different levels of abstraction. In this application, the data are sent to Hackystat services using plugins installed on data sources. The data is received by Sensor base, the root service of the application, which is responsible for persistence management and handling of configurations along with notification operations. The other services, DailyProjectData and Telemetry work via interaction with Sensorbase. These are used to compute daily, weekly, monthly and yearly abstractions of data. ProjectBrowser and TickerTape are client components used to present metrics through graphical user interfaces and post information on external applications like Nabaztag Rabbit and Twitter. The developers aim to reengineer Hackystat to serve as SaaS. It is expected to have the capability to scale for the required computing and storage resources. In this scenario, Hackystat's services are aimed to move to Amazon EC2 elastic computing and Google app engine".* [46], page.82

**Evaluation procedure.** MLSAC retrieves the base reengineering method model listing mandatory SaaS-specific method fragments (migration type II [3]) from the repository. These include method fragments such as *Isolate tenant availability*, *Isolate tenant customizability*, *Isolate tenant data*, *Isolate tenant performance*, *Handle transient faults*, *Identify incompatibilities*, *Analyze business requirements*, *Analyze migration cost*, and *Analyze migration feasibility*. The interaction forms in MLSAC enable the method engineer to tailor, e.g., adding, removing, extending, or modifying, this method regarding the characteristics of this reengineering scenario. For example, in the document of Hackystat method the task called *Requirement identification* which is to identify high-level requirements initiating the migrating Hackystat to the cloud. The task is stored as the method fragment *Analyze migration requirements* in the method. Hackystat report describes M1 level task *Identification of potential cloud hosting*, which is to list candidate cloud platforms that may address confidentiality and sensitivity requirements. Subsequently, potential incompatibilities between the legacy application and candidate cloud platforms are analyzed in a task named *Analysing applications' compatibility*. The method engineer structures and stores this fragment in MLSAC repository as the M2-level method fragment *Identify incompatibilities* in *Design* phase. The details of this task are stored in the method fragment's definition part. She can add new method fragments to the method as upcoming requirements arise during the project. For example, she defines an operationalization technique for the method fragment *Enable elasticity* in *Enable* phase. In this scenario, she uses three existing resource provision techniques based on the existing literature [47-49] (Figure 7): (i) *Reactive scaling* where developers define a set of threshold-based scaling rules for resource acquisition and release which requires a deep knowledge of the application resource utilisation patterns, (ii) *Proactive scaling* where developers use observation and prediction techniques to anticipate workload, and (iii) *Hybrid scaling* where a combination of reactive and proactive techniques are used to determine when to get a resource during a period of application execution. These techniques are then assigned to *enable elasticity*. Developers enact this base method for a scenario of migrating open-source software, named Hackystat, to two cloud platforms Amazon EC2 and Google App Engine.

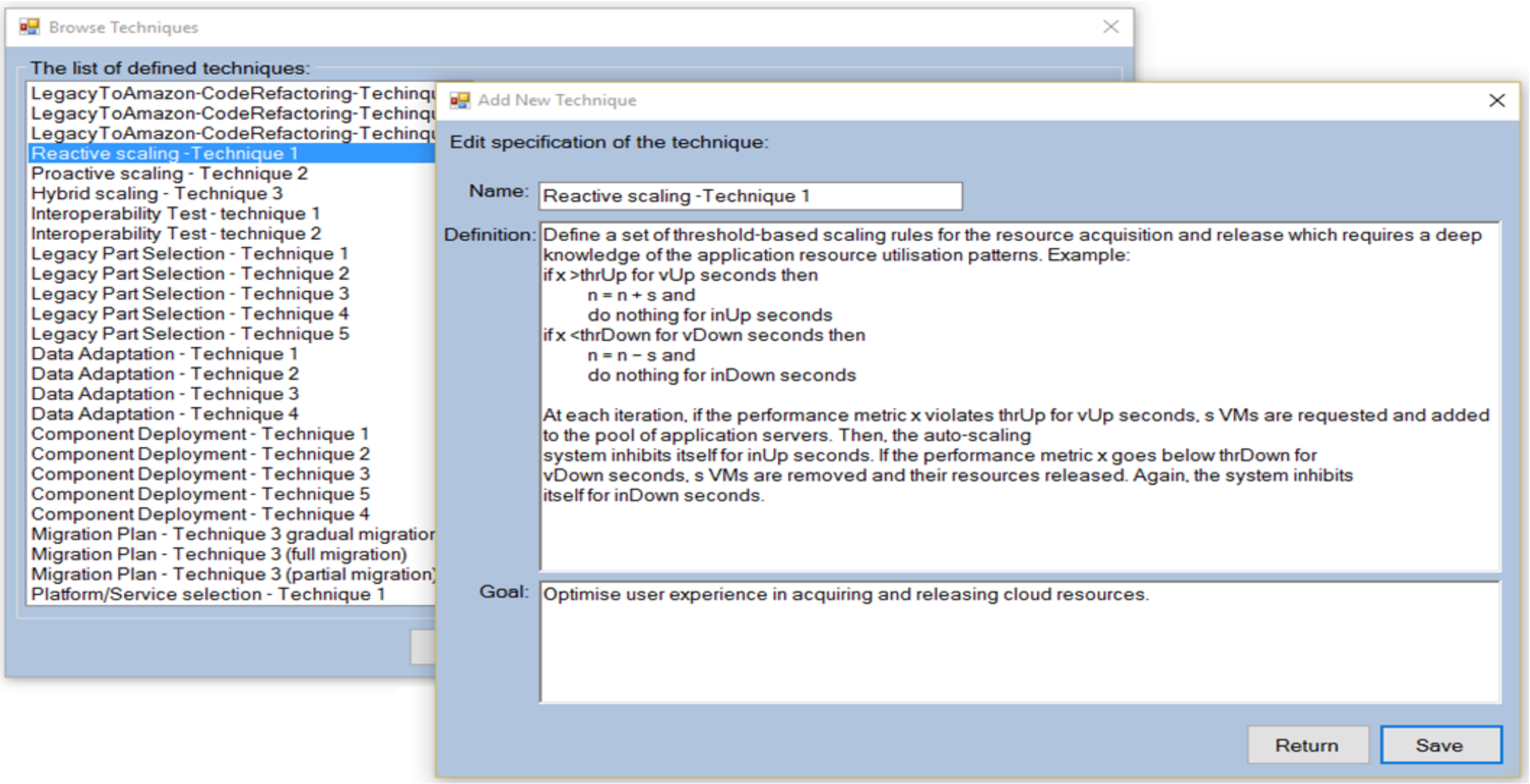

Figure 7. Defining resource scaling implementation techniques in MLSAC repository to be used in method fragments

As the result of the evaluation, the refinement was made to *Choose cloud service platform* defined in the *Design* phase to include high-level criteria for the cloud platform selection. According to Hackystat, the criteria were added to the definition of the task: budgetary constraints of a project, support of business domain of the project, and legal requirements.

## 5.3 User evaluation

We sought the opinions of two industry partners about MLSAC adherence to all the quality factors. We recruited a purposive sample [50] of interviewees to examine if MLSAC satisfies the quality factors. Two independent experts, called E1 and E2, from two different companies were selected based on the criteria (i) having real-world experience in cloud migration and (ii) speaking English fluently. The users had been involved in multiple cloud migration projects and therefore potential bias towards their evaluation of framework was not considered as a threat. The profiles of users were as follows:

- E1 was a senior .Net developer and technical lead at digital consulting firm Deloitte Digital in Sydney, Australia, with expertise in developing SaaS applications. He had been involved as a technical lead in reengineering legacy customer relationship management (CRM) applications to serve as SaaS.
- E2 was a full stack iOS engineer at Nudge group in Sydney, Australia, with expertise in services for cloud-based application development, in particular, NoSQL and Amazon Web Service (AWS). He was the technical lead in implementing a real-time and location-based social network mobile application in the online dating domain.

We organized face-to-face individual meetings; each took about 180 minutes including follow-up discussions. The research objectives, description of quality factors, and example screenshots of MLSAC were presented to them. Each expert user was asked to model the in-house reengineering method for a scenario that she/he wanted to provide for the software team to enact. The following criteria were set to select a scenario for the MLSAC evaluation: (i) having clear goals of reengineering such as the improving scalability, or performance of legacy applications, and (ii) involving with cloud-specific concerns during reengineering such as interoperability, platform selection, and server latency [3],[51]. Scenarios 1 and 2, classifying under the migration types IV/V and IV respectively, were as follows:

***Case study 1:*** *reengineering a legacy CRM application to SaaS.* The CRM application was unable to support new business requirements such as scalability for the growing number of application users (processing more than 2000 user transactions per second). SaaS version of the CRM could be a viable solution to address this critical requirement. A generic method could provide an overall road map for making CRM application SaaS-enabled.

***Case study 2:*** *migrating a real-time geosocial networking application to the cloud*. The application was recognizing the geographical zone of a registered person in the application and suggests upcoming events in that zone. The application data layer was using a relational database hosted on local servers. Over time, the relational data was found lacking in scalability since the database size was growing and search queries were becoming complicated. The application also lacked a real-time response to features such as instantaneous upload/download operations for resources used frequently. Migrating the database components to No-SQL and running the business logic components in Amazon cloud servers could improve the query performance.

Table 5. Summary of domain expert evaluation results

| Factor s | Assessment question | Evaluation results |
|---|---|---|
| Semantic | Does MLSAC repository provide necessary and relevant method fragments for representing reengineering processes to cloud platforms? | Yes, the repository provides major generic tasks and it is fully customizable so it can be extended easily based on project needs (E1).<br>Yes, an advantage of the prototype is its extensibility and customizability for different needs (E2). |
| | Have names/definitions been used in the forms been clear and helpful? | Yes. Name and definitions are generic and easy to follow. The current version [of MLSAC] is understandable enough for a technical lead to finish the reengineering process. However, the user interface/user experience can be enhanced (E1).<br>Yes (E2) |
| | Are visualizations e.g., the tree-view structure is understandable and helpful for organizing processes? | Yes, tree-view is easy to understand and helps to come up with an organized structure of the plan, but the order and relation of tasks are not very intuitive. It would be nice if I could change the order of tasks easier (e.g., drag and drop). I also noticed that the newly defined tasks or subtasks are always appended to the end of the corresponding branch and it is not currently possible to change the order (E1).<br>Yes, nevertheless, it would be good if the visualization was able to show iterative development. This notion could be realized by showing a simple icon in the tasks (E2). |
| | Is the classification of method fragments based on the migration types is correct? | Yes (E1)<br>Yes, the flexibility of the framework allows modifying the classification of the method fragments (E2). |
| Tailorability | Does MLSAC provide sufficient support of necessary parameters for process tailoring? | Yes, because the prototype provides customization support if required. However, it would be nice if the user could have access to a list of suggestive tasks classified under different domains like Mobile cloud, etc. (E1).<br>No, a hybrid process is hard to support by the current prototype (for example, both types I and V) and cannot be easily defined by the current version. As such, users need to choose a migration type that is conceptually similar to the migration type (E2). |
| | Are the defined steps in MLSAC easy to perform for process creation/configuration/maintenance/sharing? | Yes, but I would also like to have a "share with email option" instead of exporting and attaching an XML file separately. In addition, it would be very nice if I could configure to share the database of MLSAC and export data in cloud spaces used by everyone who needs to be exposed to the generated data (E1).<br>Yes, but a Web-based version of the prototype could be more efficacious (E2). |
| | Does MLSAC reduce efforts for process tailoring? | Yes, but there is a lack of support for reusable templates to be used as a starting point based on different architecture design styles which can lead to better efficiency by saving time and increasing user satisfaction (E1).<br>Yes, however, it would be great if the prototype could support pre-defined templates for different legacy system types such as finance, insurance, and e-commerce (E2). |
| | Does MLSAC provide a suitable environment for process tailoring in a given migration scenario to the cloud? | Certainly, there is room for improvement, but the main features are there (E1).<br>Yes, in comparison with other existing tools like Microsoft Project, MLSAC provides a pre-built rich repository of important items required for creating migration strategies. This feature protects users from missing some important considerations for cloud migration (E2). |
| | Does MLSAC facilitate reuse in designing bespoke reengineering processes? | Yes (E1).<br>Yes. The pre-built-in repository is helpful (E2). |
| | Is MLSAC useful for sharing reengineering processes among development teams? | Yes (E1).<br>As a suggestion, it would be good if the prototype could be Web-based with support for multiple user support. Users could simultaneously work on the method and share it. With the current version of the prototype, there is a need to multiple saves and restore the XML file of a method, which may cause inconsistency of the method content. Furthermore, it would be good if MLSAC would be able to keep track of method changes such as adding, removing, modifying tasks during the method lifetime (E2). |
| Pragmatic | Do you believe MLSAC is a practical tool for its audiences? In what ways do you think MLSAC would create value for the audience? Please explain why. | The method fragments are complete. Due to the fact that the repository is fully customizable, I believe it does provide all necessary components and I can't think about a major improvement (E1).<br>The prototype is simple and easy to use to the point of meeting major objectives of a method customization process; however, it could be further enhanced to include more features and a more professional look and feel.<br>The prototype system saves time for creating migration strategies by proving a pre-built-in repository. Visualization instead of documentation helps to a better understanding of the process. XML output can be integrated with other tracking and visualization systems (E2). |

**Evaluation procedure.** The E1 and E2 individually used MLSAC to derive their in-house methods via reusing MLSAC metamodel method fragments. In each scenario, the users could configure the method including phases, tasks, operationalization techniques, and sequences to meet the scenario requirements. For example, Figure 8 shows the corresponding conceptual representation of the method for Deloitte Digital via MLSAC method fragments. During the interviews, we used a questionnaire form (Appendix A) to capture the users' feedback in line with the quality factors. The feedback of users is presented in Table 5.

**Results.** The overall feedback from the users was positive along with suggestions for the improvements of the metamodel. Both E1 and E2 mentioned that providing such a rich repository of the method fragments with the possibility for an extension with new method fragments are excellent features of MLSAC. They believed that MLSAC positively contributes to the quality of the reengineering process as its comprehensiveness feature helps method engineers avoid missing any important method fragments for inclusion into a reengineering project. E1 highlighted that MLSAC is helpful for practitioners who may not be familiar with cloud migration concepts. E2 noted that MLSAC is a move from a text-based presentation to a well-structured one that potentially facilitates method integration with other process modeling tools. Regarding the adherence to the semantic quality factor, both experts agreed that MLSAC covers major method fragments that are incorporated into a typical reengineering scenario. They also acknowledged the clarity of names, definitions, and classifications used in MLSAC.

The users raised some deficiencies in MLSAC regarding the pragmatic quality factor, specifically to user interface design. For example, E1 suggested adding a drag and drop feature for moving method fragments among phases to give better flexibility in working with the tool. Both E1 and E2, jointly, requested adding pre-built reusable templates relevant to specific domains such as mobile cloud computing, finance, or insurance as it seems MLSAC to be generic. Such templates would facilitate method creation, increase reusability, and reduce the tailoring effort for a given domain. They also suggested adding versioning control allowing multiple users to concurrently work on the same method in a way that if a user changes the method content, then MLSAC can automatically integrate this change into new a version of the method instance. Such concurrent access to a method is not supported by MLSAC at the current stage of this research. Furthermore, E2 suggested adding warning messages when users define an illogical sequence among method fragments that is against common reengineering scenarios, i.e., knowledge source. The above suggestions, yet applicable, but are outside of the main objectives of the current research and are noted as possibilities for future works.

## 5.4 Findings

MLSAC's metamodel captures a collection of typical and reusable method fragments for incorporation into typical cloud-specific reengineering processes. The method fragments have been carefully identified from the cloud computing literature and iteratively evaluated and revised towards its purported quality factors

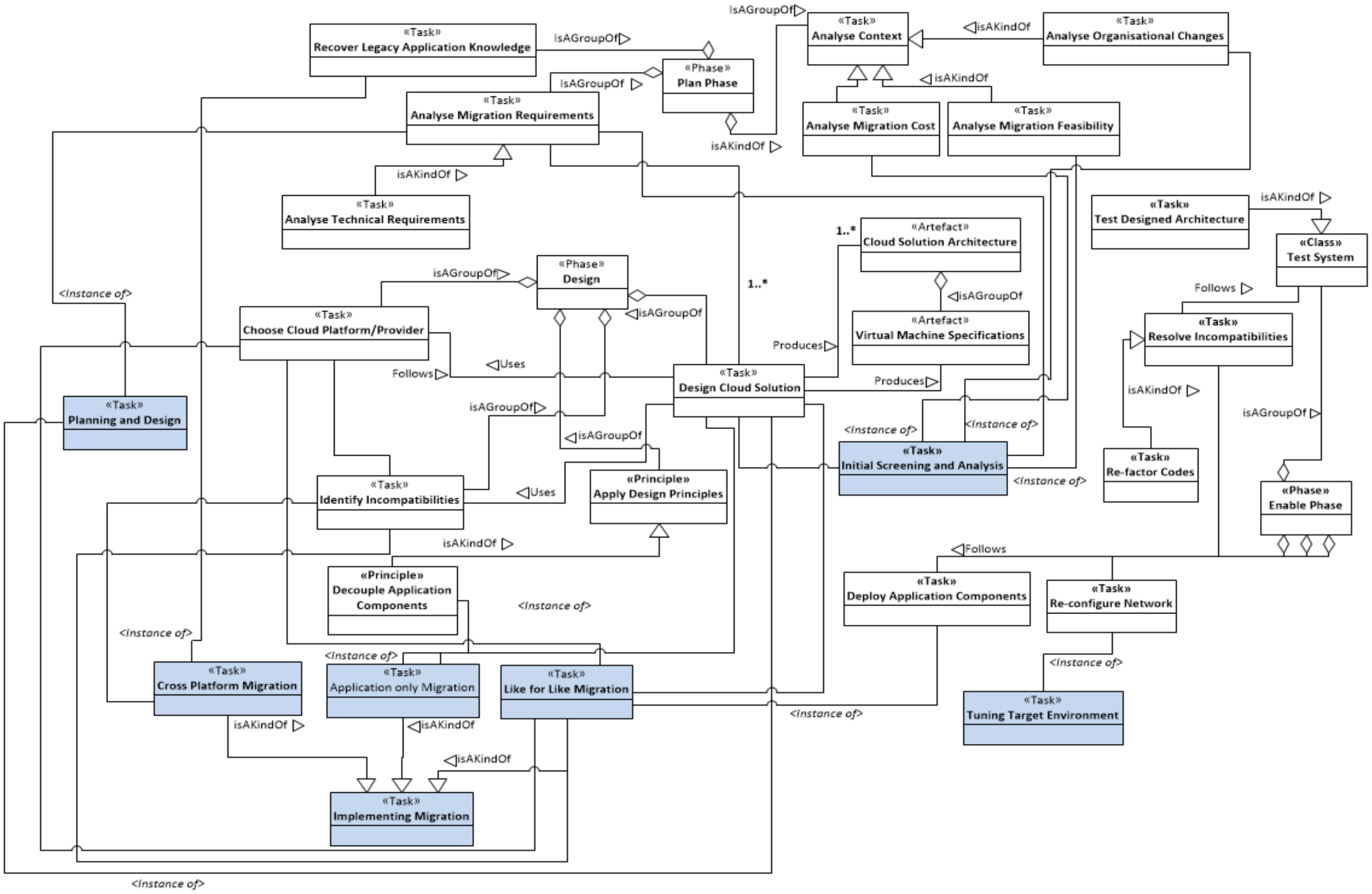

Figure 8. The representation of the corresponding reengineering process for Deloitte Digital through reusing MLSAC metamodel

. We report on these quality factors as a set of observations.

Firstly, an observation regarding the semantic quality factor is that we cannot expect to achieve complete coverage of method fragments since every reengineering process may have its own detailed technical method fragments. However, the semi-formalism and modularity of the metamodel allow refining it to new method fragments if new requirements arise. Apart from that, during the third evaluation (Section 5.3), we found that development teams may have their in-house methods for cloud migration projects. Our industry partners commonly agreed that MLSAC can be served as a good checklist and guidelines to assess the semantic quality of their in-house methods. Furthermore, an observation regarding the evaluation is that MLSAC has been more capable of representing commonalities and frequent concepts in reengineering processes rather than representing commonalities in the orders/sequences of method fragments. We believe that the definition of order for the method fragments is situation-specific, fluid, and attuned to the context of a project.

Secondly, an observation related to the tailorability quality factor is that achieving an appropriate level of abstraction for the development process description is a challenge. This is related to the granularity level chosen to represent a domain's concepts [52]. On the one hand, we had a tension to keep method fragments inclusive, generic, and applicable to represent different reengineering scenarios. On the other hand, the abstraction, inevitably, has caused the possibility of missing potential fine-granular and platform-specific method fragments in the repository. Re-structuring and storing an existing reengineering method, which is typically presented in a textual format, into MLSAC might be a cumbersome task initially as discussed in Section 5.2. However, once the method is stored, its tailoring and maintenance will be more effective as MLSAC provides modularity and separation of concerns between method design and method operationalization.

## 6 Related Work

A wide range of modeling languages has been proposed in the cloud computing literature. While they overlap, they apply several diverse modeling viewpoints to cloud-based application development. We used the taxonomy proposed by Bergmayr et al. [53] to classify the areas of concerns and capabilities of the languages and to also compare our proposed framework with the existing works. The taxonomy defines aspects of *modeling design for cloud service deployment*, *elasticity*, *cloud services*, and *application architecture*. We discarded studies related to conventional legacy application reengineering as they fail to address technical and non-technical issues specific to reengineering for the cloud. For example, reengineering methods proposed by Sneed et al. [54], Bianchi et al. [55], Stroulia et al. [56], to name a few, are too general and do not provide method fragments pertinent to cloud migration, such as the heterogeneity of cloud services and legacy applications, multitenancy, dynamic scalability, and data security [3],[51].

Reengineering to cloud platforms shares similar characteristics and challenges with other themes of legacy software reengineering to Internet-based platforms. For example, in SOA hyped organizations, migration projects aim to enable new service chains to third parties. The integration and interoperability of multiple, redundant, and dispersed data with the service providers are certainly needed. The key contribution of the work by Razavian and Lago [57] is to design a conceptual model to classify tasks involved. These include code analysis, architecture recovery, service design, and implementation to integrate legacy applications to Web Services. An advantage of our framework over this work and other similar SOA reengineering studies, e.g., [57],[58],[59],[60],[61], is that it cates for cloud-specific reengineering challenges that have been less visible in SOA. These include multi-tenancy, scalability, statelessness, multiple cloud interoperability, application licensing, legal issues, and unpredictability of cloud services. Moreover, our MLSAC framework incorporates task method fragments *identify incompatibilities* in *design phase* (Figure 2.b) and *resolve incompatibilities*, which itself includes sub-tasks *adapt data*, *refactor code*, and *develop integrator* in *Enable phase* (Figure 2.c) into addressing integration and interoperability between on-premise software systems and cloud services. The operationalization of these method fragments can be augmented by the integration techniques presented in [57],[58],[59],[60],[61]. The following further elaborates how our MLSAC framework is positioned in the literature and the way it supersedes notable related works.

### 6.1 Modeling related to cloud service deployment

Modeling enables developers to represent the target configuration and deployment of cloud applications that are a composition of cloud services. From this perspective, modeling languages are used to represent the location of services, availability zones, and storage services. Generated models are processed by tools to initiate service provisioning of computing and storage services based on deployment topology. Feature models and ontologies are a means to represent a deployment environment. For instance, computation and storage services can be captured as features of a cloud environment. MULTICLAPP [62] and the approach of Nhan et al. [63] adopt feature models to configure the target cloud environment by selecting the required cloud services. In contrast, MLSAC's objective is to provide a broad and general-purpose method model, including cloud service deployment-related method fragments, which is fundamentally at a higher level of abstraction compared to the above works. These works can be viewed as means to operationalize method fragment *Configure environment* in the *Maintain phase* of MLSAC's metamodel (Figure 2.d).

### 6.2 Modeling related to elasticity

Elasticity modeling provided by cloud platforms is used to define upper and lower bounds for service instances and elasticity rules based on which new resources are provisioned and released. CloudMIG [64] provides a modeling environment to describe elasticity rules for hardware level, e.g., CPU and storage. On the other hand, RESERVOIR-ML [65] and StratusML [66] offer a language for defining elasticity rules at a service level. Unlike CloudMIG [64], RESERVOIR-ML [65], and StratusML [66], we narrowed our view to the process lifecycle aspect. MLSAC itself defines *enable elasticity* method fragment in *Enable* phase (Figure 2.c). However, it does not provide an implementation technique for elasticity, whereas CloudMIG, RESERVOIR-ML, and StratusML define techniques to realize it. As stated earlier, the realization of method fragments goes beyond our metamodeling goal.

### 6.3 Modeling related to cloud services

The modeling can be used to represent QoS policies, constraints, and requirements such as scaling latency, 24/7 availability, and data security to be satisfied by cloud services. For example, models express terms such as *response time < 3 sec* or *Data storage is only within the Netherlands*. The application modeling in this group of studies is to ensure if service provisioning satisfies the service consumer's QoS. There are few examples in the literature supporting modeling concepts for capturing service levels using either a structured language or natural language. GENTL [63] is capable of describing QoS constraints. Using predefined stereotypes and relational operators such as *=,>, <, and key-value pairs*, MULTICLAPP [62] allows capturing QoS constraints. TOSCA [66], on the other hand, enables developers to define policies for expressing QoS that a cloud service can declare to expose. These works largely omit the method modeling perspective, which is central to our research project.

### 6.4 Modeling related to application architecture

In [67],[68] metamodels have principally been employed to ease legacy application code refactoring to enable interactions with cloud services and to facilitate interoperability across multiple cloud platforms. This is based on feature-oriented techniques for managing requirements variability and transformation techniques to instantiate an application description into multiple cloud platforms. The MULTICLAPP [62] proposes a cloud-based application development in three phases. In the application modeling phase, UML is used to represent a cloud platform-independent model of the application. The model is transformed to class skeletons and subsequently to an XML-coded deployment plan containing target cloud- information. Metamodels have also been developed to address the issue of application interoperability over multiple cloud platforms. These include CloudML [69], C3 [70], WSDL metamodel [71], Cloud-Agnostic Middleware [72], OCCI metamodel [73] and many others. For instance, the CloudML [69] proposes an extension of SOAML (Service-oriented architecture Modeling Language) to model network resources required by applications from cloud services. The CloudML engine generates script models using the JClouds APIs where they define necessary adapters to allow the application deployable across multiple cloud platforms. In MODAClouds [68], three layers of application models are defined as follows: (i) computation independent models (CIM) to represent non-functional requirements, (ii) cloud provider independent models (CPI) to include generic cloud concepts, and (iii) cloud provider model (CPM) in which details of a chosen cloud platform are added to models. In contrast to the above works, our metamodel broadens its view throughout the reengineering process instead of code level, though one can utilize the above techniques during MLSAC instantiation to a specific reengineering method.

Frey et al. [74] propose using OMG's Knowledge Discovery Metamodel to extract a utilization model of legacy application architecture, including statistical properties such as service invocation rates over time and submitted datagram sizes per request. Such models are then automatically transformed to multiple cloud platforms. The work by Zhou et al. [75] is to identify application components that their deployment on cloud servers makes business value. Its five-step approach creates an ontology of the application architecture to decompose it into candidate services. These candidate services are recommended to users for deployment and execution on cloud platforms. In another work by Hamdaqa et al., [76] a metamodel is suggested to represent main design principles, configuration rules, and semantic interpretation related to a cloud-based architecture. This facilitates a high-level architecture design of applications independent of cloud platforms. We believe these works could provide further supportive techniques for operationalization of MLSAC method fragment named *Design cloud solution* under *Design* phase (Figure 2.d).

## 7. Research Threats

As confirmed in prior studies [34],[77], the design and evaluation of metamodels for software system development is a very challenging exercise. This is due to the complex nature of metamodels, as such artifacts include many elements of phases, activities, tasks, and work-products with many different instantiation possibilities. The following describes key research threats against achieving the expected quality factors that we set in Section 4.1 and countermeasures we applied to minimize these threats.

### 7.2.1 Construct validity

The construct validity concerns the adequacy of measures used to test a concept studied [78]. The construct validity in our research is related to the evaluation of MLSAC framework adherence to the quality factors and if these factors have been clearly defined (Section 5). For example, the user evaluation in Section 5.3 was based on the questionnaire form (Appendix A). A noticeable issue related to construct validity is if the questionnaire is understandable and unambiguous for the users to evaluate MLSAC. We have minimized this threat by (i) designing questions that originated from the literature on cloud migration and method engineering and (ii) selecting qualified users with hands-on experience in legacy application reengineering cloud platforms to evaluate the framework (Section 5.3).

### 7.2.2 Internal validity

Internal validity concerns situations that may have affected the research outcome, but the researcher had not been aware of [78]. Firstly, concerning the semantic quality factor, MLSAC development has been based on the collected results from the large number of studies published on the application reengineering to the cloud where each study could have its cloud-specific method fragments [3]. Our metamodeling endeavour through use of a rigorous procedure as discussed in Section 4.2 is still circumscribed with the risk of subjective interpretation of the authors for the inclusion, exclusion, and classification of method fragments to derive the metamodel. To alleviate an imperfect analysis, we sought the active involvement of domain experts during the iterative MLSAC development and evaluations to enhance our understanding and assumptions about reengineering processes.

Secondly, as for the tailorability quality factor, the evaluation in Section 5.3 might have been undergone by the inferential capabilities of users working with the prototype system. In addition, as one of the authors of this paper has assisted the users in conducting the MLSAC evaluation steps, the evaluation results may have been biased. With these concerns in mind, we have tried our best to identify qualified experts with real-world experience in cloud migration to carry out the framework evaluation. As the users have been involved in multiple cloud migration projects and we have double-checked with them the accuracy of evaluation results, we believe the likeliness of bias is negligible.

### 7.2.3 External validity

External validity threats relate to the extent to which the research outcome can be generalized [78]. Regarding the semantic quality factor, our metamodeling effort has tended to capture commonly used method fragments grounded in the existing literature and practice. There might be some less commonly cited method fragments that are still important, however, MLSAC does not include them. We do not claim that MLSAC can represent all reengineering methods in different scenarios.

Other than that, there is no claim about the inclusivity of MLSAC's procedure for different scenarios of method tailoring. The evaluation findings summarised in Table 5 are essentially based on the users' opinions. This may limit the generalizability of the MLSAC in adhering to tailorability quality factor if MLSAC is used in further reengineering scenarios. However, we ensured that the logic of tailoring procedure is platform-independent and it can be extended to new steps. Finally, an important limitation to the generalizability of the pragmatic quality factor is originated from the fact that we have evaluated MLSAC based on our observations, the analysis of the existing methods, and opinions from two domain experts. We could attain more confidence in adherence to the pragmatic quality factor by evaluating MLSAC in more and larger cross-sectional backgrounds.

## 8 Summary and Future Work

Using the DSR approach, we presented MLSAC, a novel framework to create, maintain, share, and tailor method fits for migrating legacy applications to cloud platforms in actual practice. MLSAC provides an extensible set of method fragments, derived from the literature, in a fashion that they complemented each other to address the most critical aspects of cloud migration processes. The three-step evaluation of MLSAC shows its practical value. We do not anticipate major refinements to the methodological approach to design and evaluate MLSAC; however, we deem several future opportunities for augmenting MLSAC according to the quality factors.

Firstly, the tailorability of MLSAC can be improved by the ecosystem encompassing it. This includes a new set of guidelines to employ the metamodel. A further improvement is to define the development roles and responsibilities associated with the method fragments allowing method engineers to track the progress of the reengineering process.

Secondly, to increase the level of method reusability, we plan to enable method engineers to explore MLSAC repository to identify family-related method fragments through advanced visualization functions such as querying and filtering. Reusing a synergistic combination of individual methods or collection of method fragments can serve as a basis for creating new hybrid reengineering methods. For example, a reengineering method might be particularly designed to organize migrating legacy application database layers to Microsoft Azure SQL database. On the other hand, other methods might be suitable to be accommodated for deploying an application stack on AWS. Method fragments from both methods can be combined to create a new hybrid reengineering method to complement each other and address the overall reengineering process.

Fourthly, the evaluation of a newly created method is the last stage of a method tailoring effort [34]. This is not yet supported by our framework. It is a promising direction to provide techniques for semi-automated reasoning on a method validity. For example, this feature can uncover inconsistencies between selected and combined method fragments and identify missing important ones according to the requirements of a reengineering scenario. A higher level of automated validation can be achieved by providing consistent management rules and constraints in MLSAC.

Finally, in line with the pragmatic quality factor, MLSAC is currently limited to the tree view control, i.e., a common user interface control to visualize complex data structures, to represent the metamodel and method instances. It may raise difficulties in method design and visualization, particularly, for specifying the sequences among method elements (Figure 4.c). We plan to add new controls such as box, connector, and visual concrete syntax to enhance the usability of the framework.

### Acknowledgements

We would like to thank our industry partners who provided detailed and constructive feedback in different stages of this research. We would like to express our special thanks to anonymous reviewers whose comments improved this paper. Finally, the work of Professor John Grundy was supported by the ARC Laureate Fellowship under Grant FL190100035.

## Appendix A

The list of questions that domain experts were asked about MLSAC during the evaluation. This was organized into six steps:

*Step i*. Evaluate the quality of MLSAC repository containing relevant domain elements for representing the current in-house models.

*Step ii*. Evaluate the definitions and clarity of the method fragments.

*Step iii*. Click the Create New Method and follow the wizard steps to create the in-house method.

*Step iv*. Once the base method is created by MLSAC, browse, and analyze to check if it covers relevant fragments that would be needed for the incorporation into the defined scenario.

*Step v*. Check if the definition of method fragments, representation, and symbols are understandable.

*Step vi*. Evaluate the simplicity of MLSAC to tailor the created models. Try to modify the method by:

*Step vii.i*. Adding new method fragments (e.g., phase, task, and work-product).

*Step vi.ii*. Modifying the existing definitions of method fragments.

*Step vi.iii*. Defining realization mechanisms/techniques and assigning them to method fragments.

*Step vi.iv*. Removing method fragments that are not necessary for inclusion in the method if needed.

*Step vi.v*. Defining relationships among method fragments as required.

## REFERENCES

[1] M. Fahmideh, F. Daneshgar, G. Beydoun, and F. Rabhi, "Challenges in migrating legacy software systems to the cloud: an empirical study," *Information Systems,* vol. 67, pp. 100-113, 2017.

[2] W. K. Assunção, R. E. Lopez-Herrejon, L. Linsbauer, S. R. Vergilio, and A. J. E. S. E. Egyed, "Reengineering legacy applications into software product lines: a systematic mapping," vol. 22, no. 6, pp. 2972-3016, 2017.

[3] M. Fahmideh, F. Daneshgar, G. Low, and G. Beydoun, "Cloud migration process—A survey, evaluation framework, and open challenges," *Journal of Systems and Software,* vol. 120, pp. 31-69, 2016.

[4] M. Shuaib, A. Samad, S. Alam, and S. T. Siddiqui, "Why adopting cloud is still a challenge?—A review on issues and challenges for cloud migration in organizations," in *Ambient Communications and Computer Systems*: Springer, 2019, pp. 387-399.

[5] F. De Angelis and A. Polini, "Evaluation of Cloud Portability of legacy applications," in *2018 IEEE/ACM International Conference on Utility and Cloud Computing Companion (UCC Companion)*, 2018: IEEE, pp. 232-237.

[6] A. Menychtas *et al.*, "ARTIST Methodology and Framework: A novel approach for the migration of legacy software on the Cloud," in *Symbolic and Numeric Algorithms for Scientific Computing (SYNASC), 2013 15th International Symposium on*, 2013: IEEE, pp. 424-431.

[7] M. K. Pratt, "Why cloud migration failures happen and how to prevent them," in *Availabe at: https://searchcio.techtarget.com/feature/Cloud-migration-failures-and-how-to-prevent-them (last access 2020)*, TechTarget Ed., 2020.

[8] M. Bayern, "Organizations fail to implement basic cloud security tools," *TechrePublic,* vol. Available at: https://www.techrepublic.com/article/organizations-fail-to-implement-basic-cloud-security-tools/ (last access 2020), 2019.

[9] J. Wilson, "User Attitudes about Securing Hybrid- and Multi-Cloud Environments," *IHS Markit Technology-White Paper,* no. Available at: https://www.fortinet.com/content/dam/fortinet/assets/analyst-reports/ar-2019-ihsm-fortinet-wp-q2.pdf (last access 2020), 2020.

[10] M. Fahmideh, F. Daneshgar, and F. Rabhi, "Cloud migration: methodologies: preliminary findings," in *European Conference on Service-Oriented and Cloud Computing–CloudWays 2016 Workshop*, 2016d.

[11] R. Rai, G. Sahoo, and S. J. S. Mehfuz, "Exploring the factors influencing the cloud computing adoption: a systematic study on cloud migration," vol. 4, no. 1, p. 197, 2015.

[12] Y. Bounagui, H. Hafiddi, and A. Mezrioui, "Requirements definition for a holistic approach of cloud computing governance," in *2015 IEEE/ACS 12th International Conference of Computer Systems and Applications (AICCSA)*, 2015: IEEE, pp. 1-8.

[13] M. Fahmideh, F. Daneshgar, F. Rabhi, and G. Beydoun, "A generic cloud migration process model," *European Journal of Information Systems,* vol. 28, no. 3, pp. 233-255, 2019.

[14] M. Fahmideh, Low Graham, Ghassan Beydoun, "Conceptualising Cloud Migration Process," *Available at: https://arxiv.org/ftp/arxiv/papers/2109/2109.01757.pdf,* vol. 1, p. 0, 2016.

[15] A. Alkhalil, R. Sahandi, and D. J. I. J. o. B. I. S. John, "A decision process model to support migration to cloud computing," vol. 24, no. 1, pp. 102-126, 2017.

[16] S. Beydeda, M. Book, and V. Gruhn, *Model-driven software development*. Springer, 2005.

[17] M. Fahmideh and R. Ramsin, "Strategies for Improving MDA-Based Development Processes," in *Intelligent Systems, Modelling and Simulation (ISMS), 2010 International Conference on*, 2010: IEEE, pp. 152-157.

[18] C. Gonzalez-Perez and B. Henderson-Sellers, *Metamodelling for software engineering*. Wiley Publishing, 2008.

[19] B. Henderson-Sellers, "Bridging metamodels and ontologies in software engineering," *Journal of Systems and Software,* vol. 84, no. 2, pp. 301-313, 2011.

[20] M. Brambilla, Jordi Cabot, and Manuel Wimmer, "Model-driven software engineering in practice," vol. 3, no. 1, pp. 1-207, 2017.

[21] M. Cervera, M. Albert, V. Torres, and V. J. I. S. Pelechano, "On the usefulness and ease of use of a model-driven Method Engineering approach," vol. 50, pp. 36-50, 2015.

[22] H. Moradi, B. Zamani, and K. J. F. G. C. S. Zamanifar, "CaaSSET: A Framework for Model-Driven Development of Context as a Service," vol. 105, pp. 61-95, 2020.

[23] R. Hebig, D. E. Khelladi, and R. Bendraou, "Approaches to co-evolution of metamodels and models: A survey," *IEEE Transactions on Software Engineering,* vol. 43, no. 5, pp. 396-414, 2017.

[24] W. L. Hürsch and C. V. Lopes, "Separation of concerns," 1995.

[25] B. Fitzgerald, G. Hartnett, and K. Conboy, "Customising agile methods to software practices at Intel Shannon," *European Journal of Information Systems,* vol. 15, no. 2, pp. 200-213, 2006.

[26] F. Karlsson and P. J. Ågerfalk, "Method configuration: adapting to situational characteristics while creating reusable assets," *Information and software technology,* vol. 46, no. 9, pp. 619-633, 2004, doi: http://dx.doi.org/10.1016/j.infsof.2003.12.004.

[27] K. Peffers, T. Tuunanen, M. A. Rothenberger, and S. Chatterjee, "A design science research methodology for

information systems research," *Journal of management information systems,* vol. 24, no. 3, pp. 45-77, 2007.
[28] S. Gregor and A. R. Hevner, "Positioning and presenting design science research for maximum impact," *MIS quarterly*, vol. 37, no. 2, pp. 337-356, 2013.
[29] P. Church, H. Mueller, C. Ryan, S. V. Gogouvitis, A. Goscinski, and Z. J. J. o. C. C. Tari, "Migration of a SCADA system to IaaS clouds–a case study," vol. 6, no. 1, p. 11, 2017.
[30] M. Fahmideh, A. Ahmad, A. Behnaz, J. Grundy, and W. Susilo, "Software Engineering for Internet of Things: The Practitioner's Perspective," *IEEE Transactions on Software Engineering,* 2021.
[31] M. Fahmideh and D. Zowghi, "An exploration of IoT platform development," *Information Systems,* vol. 87, p. 101409, 2020.
[32] C. Atkinson and T. Kuhne, "Model-driven development: a metamodeling foundation," *Software, IEEE,* vol. 20, no. 5, pp. 36-41, 2003.
[33] A. G. Kleppe, J. Warmer, W. Bast, and M. Explained, "The model driven architecture: practice and promise," ed: Addison-Wesley Longman Publishing Co., Inc., Boston, MA, 2003.
[34] B. Henderson-Sellers and J. Ralyté, "Situational Method Engineering: State-of-the-Art Review," *J. UCS,* vol. 16, no. 3, pp. 424-478, 2010.
[35] L. M. Rose, D. S. Kolovos, R. F. Paige, and F. A. Polack, "Model migration with epsilon flock," in *International Conference on Theory and Practice of Model Transformations,* 2010: Springer, pp. 184-198.
[36] O. m. group, "Software process engineering metamodel specification," *Adopted Specification of the Object Management Group, Inc; Version,* vol. 1, 2006.
[37] T. Mens and P. Van Gorp, "A taxonomy of model transformation," *Electronic Notes in Theoretical Computer Science,* vol. 152, pp. 125-142, 2006.
[38] S. Sendall and W. Kozaczynski, "Model transformation the heart and soul of model-driven software development," 2003.
[39] O. I. Lindland, G. Sindre, and A. Solvberg, "Understanding quality in conceptual modeling," *Software, IEEE,* vol. 11, no. 2, pp. 42-49, 1994.
[40] M. Fahmideh, P. Jamshidi, and F. Shams, "A procedure for extracting software development process patterns," in *Computer Modeling and Simulation (EMS), 2010 Fourth UKSim European Symposium on*, 2010: IEEE, pp. 75-83.
[41] B. Kitchenham, O. Pearl Brereton, D. Budgen, M. Turner, J. Bailey, and S. Linkman, "Systematic literature reviews in software engineering – A systematic literature review," *Information and software technology,* vol. 51, no. 1, pp. 7-15, 2009, doi: http://dx.doi.org/10.1016/j.infsof.2008.09.009.
[42] T. Greenhalgh and R. Taylor, "How to read a paper: Papers that go beyond numbers (qualitative research)," *BMj,* vol. 315, no. 7110, pp. 740-743, 1997.
[43] M. Fahmideh, "Auxiliary material- the list of studies related to reengineering legacy applications to cloud," *Available at: https://www.researchgate.net/publication/353287595_Auxiliary_material-_the_list_of_studies_related_to_reengineering_legacy_applications_to_cloud*, 2021.
[44] G. Wachsmuth, "Metamodel adaptation and model co-adaptation," in *European Conference on Object-Oriented Programming*, 2007: Springer, pp. 600-624.
[45] MLSAC, *publicly available at: https://github.com/MahdiFahmideh/MLSAC*. 2019.
[46] M. A. Chauhan and M. A. Babar, "Towards Process Support for Migrating Applications to Cloud Computing," in *Cloud and Service Computing (CSC), 2012 International Conference on*, 22-24 Nov. 2012 2012, pp. 80-87, doi: 10.1109/csc.2012.20.
[47] T. Lorido-Botran, J. Miguel-Alonso, and J. A. Lozano, "A review of auto-scaling techniques for elastic applications in cloud environments," *Journal of Grid Computing,* vol. 12, no. 4, pp. 559-592, 2014.
[48] G. Galante and L. C. E. d. Bona, "A survey on cloud computing elasticity," in *Utility and Cloud Computing (UCC), 2012 IEEE Fifth International Conference on*, 2012: IEEE, pp. 263-270.
[49] T. Lorido-Botrán, J. Miguel-Alonso, and J. A. Lozano, "Auto-scaling techniques for elastic applications in cloud environments," *Department of Computer Architecture and Technology, University of Basque Country, Tech. Rep. EHU-KAT-IK-09-12,* 2012.
[50] O. C. Robinson, "Sampling in interview-based qualitative research: A theoretical and practical guide," *Qualitative research in psychology,* vol. 11, no. 1, pp. 25-41, 2014.
[51] V. Andrikopoulos, T. Binz, F. Leymann, and S. Strauch, "How to adapt applications for the Cloud environment," (in English), *Computing,* vol. 95, no. 6, pp. 493-535, 2013/06/01 2013, doi: 10.1007/s00607-012-0248-2.
[52] B. Henderson-Sellers and C. Gonzalez-Perez, "Granularity in conceptual modelling: application to metamodels," in *Conceptual Modeling–ER 2010*: Springer, 2010, pp. 219-232.
[53] A. Bergmayr *et al.*, "A systematic review of cloud modeling languages," *ACM Computing Surveys (CSUR),* vol. 51, no. 1, p. 22, 2018.
[54] H. M. Sneed, "Planning the reengineering of legacy systems," *Software, IEEE,* vol. 12, no. 1, pp. 24-34, 1995, doi: 10.1109/52.363168.
[55] A. Bianchi, D. Caivano, V. Marengo, and G. Visaggio, "Iterative reengineering of legacy systems," *Software Engineering, IEEE Transactions on,* vol. 29, no. 3, pp. 225-241, 2003, doi: 10.1109/tse.2003.1183932.
[56] E. Stroulia, M. El-Ramly, and P. Sorenson, "From legacy to web through interaction modeling," in *Software Maintenance, 2002. Proceedings. International Conference on,* 2002: IEEE, pp. 320-329.

[57] M. Razavian and P. Lago, "A lean and mean strategy: a data migration industrial study," *Journal of Software: Evolution and Process,* pp. n/a-n/a, 2013, doi: 10.1002/smr.1613.

[58] A. Umar and A. Zordan, "Reengineering for service oriented architectures: A strategic decision model for integration versus migration," *Journal of Systems and Software,* vol. 82, no. 3, pp. 448-462, 2009, doi: http://dx.doi.org/10.1016/j.jss.2008.07.047.

[59] R. Khadka, G. Reijnders, A. Saeidi, S. Jansen, and J. Hage, "A method engineering based legacy to SOA migration method," in *Software Maintenance (ICSM), 2011 27th IEEE International Conference on*, 25-30 Sept. 2011 2011, pp. 163-172, doi: 10.1109/icsm.2011.6080783.

[60] R. Khadka, B. V. Batlajery, A. M. Saeidi, S. Jansen, and J. Hage, "How do professionals perceive legacy systems and software modernization?," in *Proceedings of the 36th International Conference on Software Engineering*, 2014, pp. 36-47.

[61] K. A. Nasr, H. G. Gross, and A. van Deursen, "Realizing service migration in industry—lessons learned," *Journal of Software: Evolution and Process,* vol. 25, no. 6, pp. 639-661, 2013.

[62] J. Guillén, J. Miranda, J. M. Murillo, and C. Canal, "A UML Profile for modeling multicloud applications," in *European Conference on Service-Oriented and Cloud Computing*, 2013: Springer, pp. 180-187.

[63] V. Andrikopoulos, A. Reuter, S. G. Sáez, and F. Leymann, "A GENTL approach for cloud application topologies," in *European Conference on Service-Oriented and Cloud Computing*, 2014: Springer, pp. 148-159.

[64] S. Frey and W. Hasselbring, "The cloudmig approach: Model-based migration of software systems to cloud-optimized applications," *International Journal on Advances in Software,* vol. 4, no. 3 and 4, pp. 342-353, 2011.

[65] B. Rochwerger *et al.*, "The reservoir model and architecture for open federated cloud computing," *IBM Journal of Research and Development,* vol. 53, no. 4, pp. 4: 1-4: 11, 2009.

[66] O. Standard, "Topology and orchestration specification for cloud applications version 1.0," ed: Technical report, OASIS Standard, 2013.

[67] J. Wettinger *et al.*, "Integrating Configuration Management with Model-driven Cloud Management based on TOSCA," in *CLOSER*, 2013, pp. 437-446.

[68] D. Ardagna *et al.*, "MODAClouds: a model-driven approach for the design and execution of applications on multiple clouds," presented at the Proceedings of the 4th International Workshop on Modeling in Software Engineering, Zurich, Switzerland, 2012.

[69] E. Brandtzæg, S. Mosser, and P. Mohagheghi, "Towards CloudML, a model-based approach to provision resources in the clouds," in *8th European Conference on Modelling Foundations and Applications (ECMFA)*, 2012, pp. 18-27.

[70] I. Brandic, S. Dustdar, T. Anstett, D. Schumm, F. Leymann, and R. Konrad, "Compliant cloud computing (c3): Architecture and language support for user-driven compliance management in clouds," in *Cloud Computing (CLOUD), 2010 IEEE 3rd International Conference on*, 2010: IEEE, pp. 244-251.

[71] R. Sharma and M. Sood, "A model driven approach to cloud saas interoperability," *International Journal of Computer Applications,* vol. 30, no. 8, pp. 1-8, 2011.

[72] E. M. Maximilien, A. Ranabahu, R. Engehausen, and L. C. Anderson, "Toward cloud-agnostic middlewares," in *Proceedings of the 24th ACM SIGPLAN conference companion on Object oriented programming systems languages and applications*, 2009: ACM, pp. 619-626.

[73] P. Merle, O. Barais, J. Parpaillon, N. Plouzeau, and S. Tata, "A precise metamodel for open cloud computing interface," in *Cloud Computing (CLOUD), 2015 IEEE 8th International Conference on*, 2015: IEEE, pp. 852-859.

[74] S. Frey and W. Hasselbring, "Model-based migration of legacy software systems to scalable and resource-efficient cloud-based applications: The cloudmig approach," in *CLOUD COMPUTING 2010, The First International Conference on Cloud Computing, GRIDs, and Virtualization*, 2010, pp. 155-158.

[75] H. Zhou, H. Yang, and A. Hugill, "An ontology-based approach to reengineering enterprise software for cloud computing," in *Computer Software and Applications Conference (COMPSAC), 2010 IEEE 34th Annual*, 2010: IEEE, pp. 383-388.

[76] M. Hamdaqa, Livogiannis, T., Tahvildari, L., "A Reference Model for Developing Cloud Applications," 2011: In: Proceedings of CLOSER 2011, pp. 98--103.

[77] T. Dyba, N. B. Moe, and E. Arisholm, "Measuring software methodology usage: challenges of conceptualization and operationalization," in *2005 International Symposium on Empirical Software Engineering, 2005*, 2005: IEEE, p. 11 pp.

[78] C. Wohlin, M. Höst, and K. Henningsson, "Empirical research methods in software engineering," in *Empirical methods and studies in software engineering*: Springer, 2003, pp. 7-23.